\documentclass[aps,prd,notitlepage,showpacs,nofootinbib,superscriptaddress]{revtex4-1}
\pdfoutput=1
\usepackage[usenames,dvipsnames]{color}  %%%%%%% usenames,dvipsnames required to get BrickRed
\usepackage{graphicx} %%%%%% do not use \usepackage[dvips]{graphicx}%%%% it will creat a clash with color n brickred will not appear%%%%%

\usepackage{latexsym}
\usepackage{amsthm}
\usepackage{amsmath}
\usepackage{wrapfig}
\usepackage{amssymb}
\usepackage{cancel}
\usepackage{multirow}%\usepackage{dsfont}
\usepackage{pst-3d}                  % PSTricks 3D framed boxes
\usepackage{pst-grad}                % PSTricks gradient mode
\usepackage{pst-node}                % PSTricks nodes
\usepackage{pst-slpe}                % Improved PSTricks gradients
\usepackage{pst-plot}                %
\usepackage{pst-coil}                %
\usepackage{pst-math}                %
\usepackage{pstricks-add}            %
\usepackage{rotating}
\usepackage{geometry}
\usepackage{hhline}
\usepackage{amsmath}
\usepackage{float}
\usepackage{fancybox}

\usepackage[colorlinks=true,citecolor=darkred,urlcolor=darkred, pdfborder={0 0 0}]{hyperref}
\usepackage[normalem]{ulem}

\usepackage{braket}
\definecolor{linkcolor}{rgb}{0,0,0.5}
\allowdisplaybreaks
\allowdisplaybreaks[1]
\allowdisplaybreaks[2]

\definecolor{greenLinks}{rgb}{0, 0.6, 0}
\definecolor{blueLinks}{rgb}{0, 0, 0.6}
\definecolor{redLinks}{rgb}{0.6, 0, 0}
\definecolor{tempText}{rgb}{0.55, 0.10,0.67}
\definecolor{eprintLinks}{rgb}{0.4, 0.4, 0.4}
\definecolor{journalLinks}{rgb}{0.6, 0, 0}
\newcommand {\ignore}[1]{}
\usepackage{soul}
\definecolor{mightnightblue}{RGB}{25,25,112}

\definecolor{brown}{rgb}{0.59, 0.29, 0.0}

\definecolor{darkred}{rgb}{0.6,0,0}

\def\lsim{\mathrel{\rlap{\lower4pt\hbox{\hskip1pt$\sim$}}
    \raise1pt\hbox{$<$}}}
\def\gsim{\mathrel{\rlap{\lower4pt\hbox{\hskip1pt$\sim$}}
    \raise1pt\hbox{$>$}}}

\def\U1s{$\mathrm{U_{1}^{(a)}\otimes U_{1}^{(b)}\otimes U_{1}^{(c)}\otimes U_{1}^{(d)}\otimes U_{1}^{(e)}}$ }
\def\3211{$\mathrm{SU(3) \otimes SU(2)_L \otimes U(1)_R \otimes U(1)_{B-L}}$ }
\def\321{$\mathrm{SU(3) \otimes SU(2) \otimes U(1)}$ }
\def\422{$\mathrm{SU(4) \otimes SU(2) \otimes SU(2)_R}$ }

\newcommand{\AddrAHEP}{%
  AHEP Group, Institut de F\'{i}sica Corpuscular --
  C.S.I.C./Universitat de Val\`{e}ncia, Parc Cient\'ific de Paterna.\\
 C/ Catedr\'atico Jos\'e Beltr\'an, 2 E-46980 Paterna (Valencia) - SPAIN}

\begin{document}
\bibliographystyle{unsrt}   % needed for refs and hyperlinks
%% utphys.bst style file is also needed with this %%%%%%%%%%%%
\vspace{-2cm}
\begin{flushright}
\ \hfill\mbox{\small USTC-ICTS-19-09}\\[4mm]
\begin{minipage}{0.1\linewidth}
\normalsize
\end{minipage}
\end{flushright}
\vspace{-1cm}
\title{\boldmath  \color{BrickRed} CP Symmetries as Guiding Posts: revamping tri-bi-maximal Mixing. Part II}

\author{Peng Chen}\email{pche@mail.ustc.edu.cn}
\affiliation{College of Information Science and Engineering,Ocean University of China, Qingdao 266100, China}
\author{Salvador Centelles Chuli\'{a}}\email{salcen@ific.uv.es}
\affiliation{\AddrAHEP}
\author{Gui-Jun Ding}\email{dinggj@ustc.edu.cn}
\affiliation{Interdisciplinary Center for Theoretical Study and Department of Modern Physics, \\
University of Science and Technology of China, Hefei, Anhui 230026, China}
\author{Rahul Srivastava}\email{rahulsri@ific.uv.es}
\affiliation{\AddrAHEP}
\author{Jos\'{e} W. F. Valle}\email{valle@ific.uv.es}
\affiliation{\AddrAHEP}

\begin{abstract}
  \vspace{1cm}

In this follow up of arXiv:1812.04663 we analyze the generalized CP symmetries of the charged lepton mass matrix compatible with the complex version of the Tri-Bi-Maximal (TBM) lepton mixing pattern. These symmetries are used to ``revamp'' the simplest TBM \textit{Ansatz} in a systematic way. Our generalized patterns share some of the attractive features of the original TBM matrix and are consistent with current oscillation experiments. We also discuss their phenomenological implications both for upcoming neutrino oscillation and neutrinoless double beta decay experiments.

\end{abstract}

\maketitle

%%%%%%%%%%%%%%%%%%%%%%%%%%%%%%%%%%%%%%%%%%%%%%%%%%%%%%%%%%%%%%%%%%%%

\section{Introduction}
\label{sec:introduction}

The structure of the lepton sector and the properties of neutrinos stand out as a key missing link in particle physics, whose understanding is required for the next leap forward.
Neutrino oscillation studies have already given us a first hint for CP violation in the lepton sector~\cite{deSalas:2017kay}, though further experiments are needed to improve our measurement of leptonic CP violation~\cite{Acciarri:2015uup,Srivastava:2018ser,Nath:2018fvw}.
Moreover we need information on the elusive Majorana phases that would show up in the description of lepton number violating processes such as neutrinoless double-beta decay~\cite{Schechter:1980gr,Schechter:1980gk}, whose discovery would establish the self-conjugate nature of neutrinos.
Improving the current experimental sensitivities~\cite{KamLAND-Zen:2016pfg,GERDA:2018,MAJORANA:2008,CUORE:2018,EXO-2018,Arnold:2016bed} is, again, necessary for the next step.

It is a timely moment to make theory predictions for mixing parameters and CP violating phases charaterizing the lepton sector. The most promising tool is to appeal to symmetry considerations~\cite{Ishimori:2010au}.
As benchmarks we have ideas such as mu-tau symmetry and the Tri-Bi-Maximal (TBM) neutrino mixing~\cite{harrison:2002er}.
Thanks to the reactor measurements of non-zero $\theta_{13}$, these benchmarks can not be the final answer~\cite{Abe:2014bwa,An:2016ses,Pac:2018scx}.
However, they capture an important part of the truth, providing a valid starting point for building viable patterns of lepton mixing~\cite{Morisi:2013qna,Chatterjee:2017ilf,Rahat:2018sgs}.

Flavor symmetries can be implemented within different approaches.
For example, one can build specific theories from scratch~\cite{Babu:2002dz,Morisi:2012fg,King:2013eh}.
Alternatively, one can adopt a model-independent framework based on the imposition of residual symmetries, irrespective of how the mass matrices actually arise from first principles~\cite{Chen:2014wxa,Chen:2015siy,Chen:2016ica,Chen:2018lsv,Everett:2015oka,Everett:2016jsk,Chen:2018zbq}.
As a step in this direction one may consider complexified versions of the standard Tri-Bi-Maximal (TBM) neutrino mixing pattern.
By partially imposing generalized CP symmetries one can construct non-trivial variants of the standard TBM \textit{Ansatz} (or any other) in a systematic manner.
Depending on the type and number of preserved CP symmetries, one obtains several different mixing matrices.
Such ``revamped'' variants of the TBM \textit{Ansatz} have in general non-zero $\theta_{13}$ as well as CP violation, as currently indicated by the oscillation data. Examples of this procedure have already been given in~\cite{Chen:2018eou,Chen:2018zbq}. The prospects for probing various neutrino mixing scenarios in present and future oscillation experiments have been discussed extensively in the literature~\cite{Pasquini:2016kwk,Srivastava:2017sno,Chatterjee:2017xkb,Chatterjee:2017ilf,Nath:2018xkz,Chakraborty:2018dew, Pasquini:2018udd, Nath:2018fvw, Ding:2019vvi}.

This paper is a follow-up of Ref.~\cite{Chen:2018zbq}.
Throughout this work we will adopt the basis in which the neutrino mass matrix is diagonal and therefore the leptonic mixing matrix  $U_{lep} = U_{cl}^\dagger$, where $U_{cl}$ is the matrix which diagonalizes the charged lepton mass matrix.
Here we take the famous \textit{Ansatz} of TBM mixing~\cite{harrison:2002er} in the charged lepton sector as a starting point, i.e. $U_{cl} = U_{TBM}^\dagger \to U_{lep} = U_{cl}^\dagger = U_{TBM}$ and seek for solutions which satisfy only a partial symmetry with respect to the full TBM symmetries. The paper is structured as follows. We start by briefly discussing remnant generalized CP symmetries in Section~\ref{sec:preliminaries} and the CP and flavour symmetries of the $U_{TBM}^\dagger$ mixing matrix in Section~\ref{sec:TBM}. We will then focus in the cases in which the mass matrix satisfies two CP symmetries in Section~\ref{sec:two-cl} and one CP symmetry in Section~\ref{sec:one-cl}.
We also discuss the phenomenological predictions from these matrices and give a brief sum-up discussion at the end.

%%%%%%%%%%%%%%%%%%%%%%%%%%%%%%%%%%%%%%%%%%%%%%%%%%%%%%%%%%%%%%%%%%%%%%%%
\section{Preliminaries}%%%%%%%%%%%%%%%%%%%%%%%%%%%%%%%%%%%%%%%%%%%
\label{sec:preliminaries}%%%%%%%%%%%%%%%%%%%%%%%%%%%%%%%%%%%%%%%%%%%%%%%
%%%%%%%%%%%%%%%%%%%%%%%%%%%%%%%%%%%%%%%%%%%%%%%%%%%%%%%%%%%%%%%%%%%%%%%%

In this section, we shall briefly review the remnant generalized CP symmetry and flavor symmetry of the charged lepton sector.
We assume that the neutrinos are Majorana particles. The neutrino and charged lepton mass term can be written as
\begin{eqnarray}
{\cal L}_{\rm mass}=-\overline{l}_R m_l l_{L}+\frac{1}{2}\nu_L^T{\cal C}^Tm_\nu\nu_L+h.c.,
\end{eqnarray}
where ${\cal C}$ is the charge-conjugation matrix, $l_L$ and $l_R$ are the three generations of the left and right-handed charged lepton fields respectively.
In this paper we work in the neutrino diagonal basis without loss of generality. As a consequence, $m_\nu$ is the diagonal neutrino mass matrix and $\nu_L$ stands for the three generation left-handed neutrino mass eigenstates.

A generalized CP symmetry combines the canonical CP transformation with a flavor symmetry.
Under the action of a generalized CP transformation, a generic fermion multiplet field $\psi$ transforms as
\begin{eqnarray}
\psi \to iX\gamma^0{\cal C}\overline{\psi}^T\,,
 \label{eq:defcp}
\end{eqnarray}
where $X$ is a unitary symmetric matrix in the flavour space which characterizes the generalized CP transformation.

Notice that for the conventional CP transformation $X$ is just the identity matrix and does not mix the flavours.
Studying the remnant generalized symmetries, i.e., the surviving generalized CP symmetries after spontaneous symmetry breaking, provides a powerful method to study the mixing pattern of leptons. Given a mixing pattern one can extract all the remnant CP symmetries of the corresponding lepton mass matrices~\cite{Chen:2015nha,Chen:2015siy}. On the other hand, one can also invert the procedure and extract the possible mixing matrices (up to some freedom) that respect one or more generalized CP symmetries. In this work we will study some aspects of such generalized CP transformations acting on the charged lepton sector.

We denote the charged lepton mass matrix as $m_{cl}$, and the hermitian mass matrix $M^2 \equiv m_{cl}^\dagger m_{cl}$ can be diagonalized by a unitary transformation $U_{cl}$,
\begin{eqnarray}
U_{cl}^\dagger M^2 U_{cl} = \text{diag}(m_e ^2, m_\mu^2, m_\tau^2)\,,
\end{eqnarray}
where $m_e$, $m_\tau$ and $m_\tau$ are the charged lepton masses.
If the charged lepton mass term is invariant under the generalized CP transformation of Eq.~\eqref{eq:defcp}, $X$ and $M^2$ should satisfy the following relation
\begin{eqnarray}
 X^\dagger M^2 X = M^{2*}\,.
 \label{eq:cptrans}
\end{eqnarray}
If this is the case, we say that $X$ is a remnant CP symmetry of the mass matrix $M^2$. As shown in several previous works ~\cite{Chen:2014wxa, Chen:2015siy, Chen:2016ica}~\cite{Chen:2018zbq}, $X$ can be written in terms of $U_{cl}$ as
\begin{eqnarray}
X = U_{cl}~\text{diag}(e^{i \delta_1},e^{i \delta_2},e^{i \delta_3})~U_{cl}^T\,,
\label{eq:cpsymcharged}
\end{eqnarray}
where the $\delta_i$ are arbitrary real parameters which label the CP transformation. In other words, given a mass matrix one can extract an infinite set of $X$ matrices satisfying Eq.~\eqref{eq:cptrans} labeled by the three real parameters $\delta_1$, $\delta_2$ and $\delta_3$. Eq.~\eqref{eq:cpsymcharged} allows us to build all the possible generalized CP symmetries from a given mass matrix or mixing pattern.

As stated before, one can also invert the logic and construct the mixing matrix given a remnant CP symmetry matrix $X$.
If the squared mass matrix $M^2$ satisfies Eq.~\eqref{eq:cptrans} it can be shown~\cite{Chen:2014wxa, Chen:2015siy, Chen:2016ica}~\cite{Chen:2018zbq} that, after Takagi decomposing the symmetric CP transformation as $X =\Sigma~ \Sigma^T$, the resulting relation between $\Sigma$ and $U_{cl}$ is
\begin{eqnarray}
 U_{cl} = \Sigma O_{3\times 3}^T Q^{-1/2}\,,
\end{eqnarray}
where $O_{3\times 3}$ is a generic real orthogonal matrix and $Q$ is a general unitary diagonal matrix. If $U_{\nu}$ is the unitary matrix that diagonalizes the neutrino mass matrix,
then the lepton mixing matrix is given as
\begin{eqnarray}
U_{lep} = U_{cl}^\dagger U_\nu\,.
\end{eqnarray}

As mentioned above, we shall work in the neutrino mass diagonal basis throughout this paper. Hence $U_\nu$ is a unit matrix and the lepton mixing matrix can be written as $U_{lep}=U_{cl}^\dagger$.

We can also follow a similar procedure to obtain the flavour symmetries of the mass matrix $M^2$. We define a remnant flavour transformation as
\begin{eqnarray}
\psi \to G \psi\,,
\end{eqnarray}
where again $G$ is a $3\times3$ unitary matrix acting in flavour space. If this transformation leaves the squared mass matrix invariant, then we say that $G$ is a residual flavour symmetry of the mass matrix and it satisfies
\begin{eqnarray}
G^\dagger M^2G=M^2\,.
\label{eq:flavtrans}
\end{eqnarray}
We can see from Eq.~\eqref{eq:flavtrans} that the flavour symmetries are all the matrices $G$ which commute with $M^2$ and therefore they share a common basis of eigenvectors.
Since the eigenvalues of $M^2$ are the squared charged lepton masses, which are non-degenerate, in the basis in which $M^2$ is diagonal $G$ will also be diagonal.
Therefore both $G$ and $M^2$ are diagonalized by the same unitary matrix $U_{cl}$, and we can build its explicit form:
\begin{eqnarray}
 U_{cl}^\dagger ~ G ~ U_{cl} = \text{diag}(e^{i \alpha},e^{i \beta},e^{i \gamma}) \to G = U_{cl} ~ \text{diag}(e^{i \alpha},e^{i \beta},e^{i \gamma}) ~ U_{cl}^\dagger\,,
 \label{eq:Gform}
\end{eqnarray}
where we used the fact that the eigenvalues of a unitary matrix must have norm $1$, so we write them as $e^{i \alpha}$, $e^{i \beta}$ and $e^{i \gamma}$, with $\alpha$, $\beta$ and $\gamma$ real.
As a side remark that will become important in Section~\ref{sec:two-cl}, note that the previous argument implies that if we impose a flavour symmetry with non-degenerate eigenvalues then the mixing matrix would be uniquely determined up to permutation of its column vectors. This holds since, if the eigenvalues of $G$ are $(e^{i \alpha},e^{i \beta},e^{i \gamma})$ with $\alpha \neq \beta \neq \gamma$,   then both $M^2$ and $G$ are diagonalized by the same unitary matrix. \\[-.2cm]

However, the above argument does not hold if the eigenvalues of $G$ are at least partially degenerate, since in this case in a basis in which $G$ is diagonal, the matrix $M^2$ may not be diagonal. There will be a subspace generated by the eigenvectors associated to the degenerate eigenvalues which won't be diagonal in general. We will exploit this feature in Section~\ref{sec:two-cl}.\\

As a final comment, let us mention that there is a relation between flavour symmetries and CP symmetries. If two CP symmetries $X_1 = U_{cl} ~ \text{diag}(e^{i \delta_{1}},e^{i \delta_{2}},e^{i \delta_{3}}) ~ U_{cl}^T$ and $X_2 = U_{cl} ~ \text{diag}(e^{i \delta_{4}},e^{i \delta_{5}},e^{i \delta_{6}}) ~ U_{cl}^T$ are preserved by the charged lepton sector, then a flavor symmetry can be induced by successively performing two CP transformations
\begin{equation}
G = X_1 X_2^* = U_{cl} ~ \text{diag}(e^{i \alpha},e^{i \beta},e^{i \gamma}) ~ U_{cl}^\dagger\,,
\end{equation}
with $\alpha=\delta_{1}-\delta_{4}$, $\beta = \delta_{2}-\delta_{5}$ and $\gamma = \delta_{3}-\delta_{6}$. In other words, applying two CP symmetries automatically implies the existence of a flavour symmetry.

%%%%%%%%%%%%%%%%%%%%%%%%%%%%%%%%%%%%%%%%%%%%%%%%%%%%%%%%%%%%%%%%%%%%%%%

\section{CP and flavour symmetries of Tri-bimaximal mixing }
\label{sec:TBM}

The goal of this paper is to modify the Tri-Bi-Maximal mixing (TBM) pattern based on the charged lepton CP symmetries. In this section we will use the results obtained in section~\ref{sec:preliminaries} to extract the CP and flavour symmetries of the celebrated TBM
\textit{Ansatz}~\cite{harrison:2002er}. This will be useful in the following sections, in which we will start from a charged lepton mass matrix satisfying the full TBM symmetry, and add a perturbation term which will satisfy only a partial symmetry.\\[-.2cm]

The standard TBM mixing pattern~\cite{harrison:2002er} is the \textit{Ansatz} in which the three mixing angles take the following values
\begin{eqnarray}
 \sin \theta_{12} =  \frac{1}{\sqrt{3}}, \hspace{2cm} \theta_{13} = 0, \hspace{2cm} \theta_{23} = \pi/4.
\end{eqnarray}
The vanishing of one of the mixing angles, in our case $\theta_{13}$, implies that the Dirac CP phase $\delta_{CP}$ of the lepton mixing matrix is unphysical~\cite{Schechter:1979bn}. Assuming zero Majorana phases, we can write the ``real TBM mixing'' as,
\begin{eqnarray}
 U_{rTBM} = U_{23}(\pi/4, 0) ~ U_{12}(\arcsin(1/\sqrt{3}, 0) =   \left(
\begin{array}{ccc}
\sqrt{\frac{2}{3}}               & \frac{1}{\sqrt{3}}           & 0   \\
-\frac{1}{\sqrt{6}}              & \frac{1}{\sqrt{3}}           & \frac{1}{\sqrt{2}} \\
\frac{1}{\sqrt{6}}               &  -\frac{1}{\sqrt{3}}         & \frac{1}{\sqrt{2}} \\
\end{array}
\right)
\label{eq:rtbm}
\end{eqnarray}
where $U_{ij}(\theta, \phi)$ is a complex rotation in the $(ij)-$plane of angle $\theta$ and phase $\phi$. For example,
\begin{eqnarray}
U_{12}(\theta,\phi) =
\left(
\begin{matrix}
 \cos{\theta} & \sin{\theta} e^{-i \phi} & 0   \\
-\sin{\theta} e^{i \phi} & \cos{\theta} & 0   \\
0 & 0 & 1
\end{matrix}
 \right ).
\end{eqnarray}
We can generalize this \textit{Ansatz} so as to include non-zero Majorana phases, thus defining the ``complex TBM mixing'' pattern as
\begin{eqnarray}
U_{cTBM} = U_{23}(\pi/4, \sigma) ~ U_{12}(\arcsin(1/\sqrt{3}),\rho)=\left(\begin{array}{ccc}
\sqrt{\frac{2}{3}}                       & \frac{ e^{-i \rho } }{\sqrt{3}}     & 0                                \\
-\frac{e^{i \rho }}{\sqrt{6}}            & \frac{1}{\sqrt{3}}                  & \frac{e^{-i\sigma}}{\sqrt{2}}    \\
\frac{e^{i (\rho + \sigma)}}{\sqrt{6}}   &  -\frac{e^{i\sigma}}{\sqrt{3}}      & \frac{1}{\sqrt{2}}               \\
\end{array}
\right)\,.
\label{eq:ctbm}
\end{eqnarray}
Note that, apart from the choice $(\rho, \sigma) = (0, 0)$, there are other choices for the phases $\rho$ and $\sigma$ which also lead to real mixing patterns: $(\rho, \sigma)=~(\pi, 0),~(0, \pi)$ and $(\pi, \pi)$.

If the lepton mixing matrix is the complex TBM matrix $U_{lep} = U_{cTBM}$, then the charged lepton mixing matrix will be $U_{cl}= U_{cTBM}^\dagger$ in the neutrino diagonal basis.
We can explicitly build the charged lepton mass matrix diagonalized by $U_{cTBM}^\dagger$ as
\begin{eqnarray}
\label{eq:McTBM}
 M^2_{cTBM} & = &\frac{m_e^2}{3} \left(
\begin{array}{ccc}
 2 & \sqrt{2} e^{-i \rho } & 0 \\
 \sqrt{2} e^{i \rho } & 1 & 0 \\
 0 & 0 & 0 \\
\end{array}
\right)
+
\frac{m_\mu^2}{6} \left(
\begin{array}{ccc}
 1 & -\sqrt{2} e^{-i \rho } & -\sqrt{3} e^{-i (\rho +\sigma )} \\
 -\sqrt{2} e^{i \rho } & 2 & \sqrt{6} e^{-i \sigma } \\
 -\sqrt{3} e^{i (\rho +\sigma )} & \sqrt{6} e^{i \sigma } & 3 \\
\end{array}
\right)
 \\ \nonumber
& + & \frac{m_\tau^2}{6} \left(
\begin{array}{ccc}
 1 & -\sqrt{2} e^{-i \rho } & \sqrt{3} e^{-i (\rho +\sigma )} \\
 -\sqrt{2} e^{i \rho } & 2 & -\sqrt{6} e^{-i \sigma } \\
 \sqrt{3} e^{i (\rho +\sigma )} & -\sqrt{6} e^{i \sigma } & 3 \\
\end{array}
\right)
\end{eqnarray}
Using Eq.~\eqref{eq:cpsymcharged} we can easily extract the CP symmetries of the matrix $M^2$, i.e. all the matrices $X$ that satisfy Eq.~\eqref{eq:cptrans},
\begin{eqnarray}
\nonumber  X & = & U_{cTBM}^\dagger ~ \text{diag}(e^{i\delta_1}, e^{i \delta_2}, e^{i \delta_3}) ~ U_{cTBM}^*\\
\nonumber& =&
  \frac{e^{i \delta_1}}{3} \left(
\begin{array}{ccc}
 2 & \sqrt{2} e^{i \rho } & 0 \\
 \sqrt{2} e^{i \rho } & e^{2 i \rho } & 0 \\
 0 & 0 & 0 \\
\end{array}
\right)+ \frac{e^{i \delta_2}}{6} \left(
\begin{array}{ccc}
 e^{-2 i \rho } & -\sqrt{2} e^{-i \rho } & -\sqrt{3} e^{-i (\rho -\sigma )} \\
 -\sqrt{2} e^{-i \rho } & 2 & \sqrt{6} e^{i \sigma } \\
 -\sqrt{3} e^{-i (\rho -\sigma )} & \sqrt{6} e^{i \sigma } & 3 e^{2 i \sigma } \\
\end{array}
\right)\\
\label{eq:Xctbm}&&+ \frac{e^{i \delta_3}}{6} \left(
\begin{array}{ccc}
 e^{-2 i (\rho +\sigma )} & -\sqrt{2} e^{-i (\rho +2 \sigma )} & \sqrt{3} e^{-i (\rho +\sigma )} \\
 -\sqrt{2} e^{-i (\rho +2 \sigma )} & 2 e^{-2 i \sigma } & -\sqrt{6} e^{-i \sigma } \\
 \sqrt{3} e^{-i (\rho +\sigma )} & -\sqrt{6} e^{-i \sigma } & 3 \\
\end{array}
\right)\,.
\end{eqnarray}
The CP symmetries which correspond to the real TBM limit in Eq.~\eqref{eq:rtbm} can be constructed from Eq.~\eqref{eq:Xctbm} just by going to the limit $\rho \to 0$ and $\sigma \to 0$. In this case the CP matrix takes the simpler form
\begin{eqnarray}
\nonumber X &=&
 U_{rTBM}^\dagger ~\text{diag}(e^{i\delta_1}, e^{i\delta_2}, e^{i\delta_3}) ~ U_{rTBM}^* \\
 &=&\frac{e^{i \delta_1}}{3} \left(
\begin{array}{ccc}
 2 & \sqrt{2} & 0 \\
 \sqrt{2} & 1 & 0 \\
 0 & 0 & 0 \\
\end{array}
\right) \, + \,
\frac{e^{i \delta_2}}{6} \left(
\begin{array}{ccc}
 1 & -\sqrt{2} & -\sqrt{3} \\
 -\sqrt{2} & 2 & \sqrt{6} \\
 -\sqrt{3} & \sqrt{6} & 3 \\
\end{array}
\right) \, + \,
\frac{e^{i \delta_3}}{6} \left(
\begin{array}{ccc}
 1 & -\sqrt{2} & \sqrt{3} \\
 -\sqrt{2} & 2 & -\sqrt{6} \\
 \sqrt{3} & -\sqrt{6} & 3 \\
\end{array}
\right)\,.
\label{eq:Xrtbm}
\end{eqnarray}
We now turn to the residual flavour symmetries of the mixing matrix $U_{cTBM}^\dagger$. Again we will extract all the matrices $G$, labeled by the three real parameters $\alpha$, $\beta$ and $\gamma$, that satisfy Eq.~\eqref{eq:flavtrans}. Using Eq.~\eqref{eq:Gform}, we find that these matrices take the form
\begin{eqnarray}
 G &=& \frac{e^{i \alpha}}{3} \left(
\begin{array}{ccc}
 2 & \sqrt{2} e^{-i \rho } & 0 \\
 \sqrt{2} e^{i \rho } & 1 & 0 \\
 0 & 0 & 0 \\
\end{array}
\right) +
\frac{e^{i \beta}}{6} \left(
\begin{array}{ccc}
 1 & -\sqrt{2} e^{-i \rho } & -\sqrt{3} e^{-i (\rho +\sigma )} \\
 -\sqrt{2} e^{i \rho } & 2 & \sqrt{6} e^{-i \sigma } \\
 -\sqrt{3} e^{i (\rho +\sigma )} & \sqrt{6} e^{i \sigma } & 3 \\
\end{array}
\right)  \\ \nonumber
& + & \frac{e^{i \gamma}}{6} \left(
\begin{array}{ccc}
 1 & -\sqrt{2} e^{-i \rho } & \sqrt{3} e^{-i (\rho +\sigma )} \\
 -\sqrt{2} e^{i \rho } & 2 & -\sqrt{6} e^{-i \sigma } \\
 \sqrt{3} e^{i (\rho +\sigma )} & -\sqrt{6} e^{i \sigma } & 3 \\
\end{array}
\right)\,.
\label{eq:GcTBM}
\end{eqnarray}
As before, we can recover the real TBM limit just by going to the limit $\rho \to 0$ and $\sigma
\to 0$.
\begin{eqnarray}
 G = \frac{e^{i \alpha}}{3} \left(
\begin{array}{ccc}
 2 & \sqrt{2} & 0 \\
 \sqrt{2} & 1 & 0 \\
 0 & 0 & 0 \\
\end{array}
\right) +
\frac{e^{i \beta}}{6} \left(
\begin{array}{ccc}
 1 & -\sqrt{2} & -\sqrt{3} \\
 -\sqrt{2} & 2 & \sqrt{6} \\
 -\sqrt{3} & \sqrt{6} & 3 \\
\end{array}
\right) +
\frac{e^{i \gamma}}{6} \left(
\begin{array}{ccc}
 1 & -\sqrt{2} & \sqrt{3} \\
 -\sqrt{2} & 2 & -\sqrt{6} \\
 \sqrt{3} & -\sqrt{6} & 3 \\
\end{array}
\right)\,.
\label{eq:GrTBM}
\end{eqnarray}
Note that if the mass matrix is real then $M^2 = M^{2*}$ and there is no difference between Eq.~\eqref{eq:cptrans} and Eq.~\eqref{eq:flavtrans}. This is why in the real TBM case the flavour symmetry matrices are identical to the CP symmetry matrices. This statement not only applies to the case $(\rho, \sigma) = (0, 0)$, which we are calling `real TBM', but also to $(\rho, \sigma) = (0, \pi)$, $(\rho, \sigma) = (\pi, 0)$ and $(\rho, \sigma) = (\pi, \pi)$.

Although the TBM \textit{Ansatz} provides an interesting starting point, note that neither the real nor the complex variants of the TBM mixing are viable lepton mixing patterns. Indeed, recent reactor measurements~\cite{Abe:2014bwa,An:2016ses,Pac:2018scx} have established that $\theta_{13}$ is non-zero with high significance. In the same spirit as \cite{Chen:2018zbq} here we show that, starting from the cTBM matrix in the charged lepton sector, and using the generalized CP symmetries, one can systematically construct and analyze realistic neutrino mixing matrices with non-zero reactor angle. An appealing feature of this method is that the resulting mixing patterns will share many properties with the original TBM \textit{Ansatz}, while avoiding the unwanted $\theta_{13} = 0$ prediction.

As a starting point we will assume neutrinos to be Majorana-type and will work in a basis in which neutrinos are diagonal. We will then start with the complex TBM matrix of Eq.~\eqref{eq:ctbm}. The real TBM matrix can always be obtained from it by simply taking the limit $\rho, \sigma \to 0$. In what follows we will take this limit at various stages of our discussion.

%%%%%%%%%%%%%%%%%%%%%%%%%%%%%%%%%%%%%%%%%%%%%%%%%%%%%%%%%%%%%%%%%%%%%%%%%%%%%%%%%%%%%%%%%%%%%%%%%%%%%%%%%%%%%%%%%%%%%

\section{Charged lepton mass matrix conserving two CP symmetries}
\label{sec:two-cl}

%%%%%%%%%%%%%%%%%%%%%%%%%%%%%%%%%%%%%%%%%%%%%%%%%%%%%%%%%%%%%%%%%%%%%%%%%%%%%%%%%%%%%%%%%%%%%%%%%%%%%%%%%%%%%%%%%%%%%%

We will start our analysis by taking as a starting point a charged lepton mass matrix diagonalized by $U_{cTBM}^{{\dagger}}$, which will be of the form shown in Eq.~\eqref{eq:McTBM}. We will then add a small perturbation which only preserves two remnant CP symmetries, and study the resulting mixing pattern, namely,
\begin{eqnarray}
M^2=M^2_{cTBM}+\delta M^2~,
\end{eqnarray}
where $\delta M ^2$ is the perturbation matrix and is therefore expected to be small.
As explained in the previous section, a CP symmetry in the charged lepton sector compatible with $U_{cTBM}$ will be of the form shown in Eq.~\eqref{eq:Xctbm} and will satisfy Eq.~\eqref{eq:cptrans}. We impose, in the perturbation term, two such symmetries given by
\begin{equation}
X_1=U_{cTBM}^\dagger~\text{diag}(e^{i\delta_1},e^{i\delta_2},e^{i\delta_3})~U_{cTBM}^\ast\,,\qquad
X_2=U_{cTBM}^\dagger~\text{diag}(e^{i\delta_4},e^{i\delta_5},e^{i\delta_6})~U_{cTBM}^\ast\,,
\end{equation}
where in general the $\delta_i$ can be different. As explained in section~\ref{sec:preliminaries}, two CP symmetries generate automatically a flavour symmetry $G_l=X_1 X_2^*$ satisfying $G_l^\dagger M^2 G_l= M^2$ and given by
\begin{equation}
G_l=U_{cTBM}^\dagger~\text{diag}(e^{i\alpha},e^{i\beta},e^{i\gamma})~U_{cTBM}\,,
\label{Tldef}
\end{equation}
with $\alpha=\delta_1-\delta_4$, $\beta=\delta_2-\delta_5$, and $\gamma=\delta_3-\delta_6$.  It is clear that
\begin{eqnarray}
U_{cTBM}~G_l~U_{cTBM}^\dagger = \text{diag}(e^{i\alpha},e^{i\beta},e^{i\gamma})\,.
\end{eqnarray}
One sees that $G_l$ is diagonalized by $U_{cTBM}^\dagger$ and its eigenvalues are $e^{i\alpha}$, $e^{i\beta}$, $e^{i\gamma}$. As shown in section~\ref{sec:preliminaries}, when $\alpha \neq \beta \neq \gamma$ the eigenvalues of $G_l$ are non-degenerate and $U_{cTBM}^\dagger$ will diagonalize both $G_l$ and $M^2$.
In this case the lepton mixing matrix would be $U_{lep}=U_{cl}^\dagger U_\nu = U_{cTBM}$ with $\theta_{13} = 0$, hence inconsistent with experiment. Here we study the particular scenarios in which $G_l$ is partially degenerate, i.e.
\begin{equation}
G_l=U_{cTBM}^\dagger ~P_l~ \text{diag}(e^{i\alpha},e^{i\alpha},e^{i\beta})~P_l^T ~U_{cTBM}\,,
\end{equation}
where $\alpha\neq\beta$ and $P_l$ is a permutation matrix which parametrizes the three possible orderings, $(e^{i\alpha}, e^{i\alpha}, e^{i\beta})$, $(e^{i\alpha}, e^{i\beta}, e^{i\alpha})$ or $(e^{i\beta}, e^{i\alpha}, e^{i\alpha})$. These permutation matrices can be given as:
\begin{eqnarray}
 P_{123}=\left(\begin{array}{ccc}
 1 & 0 & 0 \\
 0 & 1 & 0 \\
 0 & 0 & 1
\end{array}\right)\,, \qquad
P_{231}=\left(\begin{array}{ccc}
 0 & 1 & 0 \\
 0 & 0 & 1 \\
 1 & 0 & 0
\end{array}\right)\,, \qquad
P_{312}=\left(\begin{array}{ccc}
 0 & 0 & 1 \\
 1 & 0 & 0 \\
 0 & 1 & 0
\end{array}\right)\,,
\end{eqnarray}
and correspond to three possible different cases respectively, as follows:
\begin{eqnarray}
\text{case } \alpha \alpha \beta: \quad P_l=P_{123} \to G_l=U_{cTBM}^\dagger ~\text{diag}(e^{i\alpha},e^{i\alpha},e^{i\beta}) ~U_{cTBM}\,, \\
\text{case } \alpha \beta \alpha: \quad P_l=P_{231} \to G_l=U_{cTBM}^\dagger ~\text{diag}(e^{i\alpha},e^{i\beta},e^{i\alpha}) ~U_{cTBM}\,, \\
\text{case } \beta \alpha \alpha: \quad P_l=P_{312} \to G_l=U_{cTBM}^\dagger ~\text{diag}(e^{i\beta},e^{i\alpha},e^{i\alpha}) ~U_{cTBM}\,.
\end{eqnarray}
%% %
In the following, we will consider the cases $\alpha \alpha \beta$ and $\alpha \beta \alpha$.
We discard the third scenario, the $\beta \alpha \alpha$ case, as it leads to the prediction $\theta_{13}=0$ and is therefore not viable.

%%%%%%%%%%%%%%%%%%%%%%%%%%%%%%%%%%%%%%%%%%%%%%%%%%%%%%%%%%%%%%%%%%%%%%%%%%%%%%%%%%%%%%%%%%%%%%%%%%%%%%%%%%%%%%%%
\subsubsection{Case $\alpha \alpha \beta$}
\label{sec:p123-Cp213}
%%%%%%%%%%%%%%%%%%%%%%%%%%%%%%%%%%%%%%%%%%%%%%%%%%%%%%%%%%%%%%%%%%%%%%%%%%%%%%%%%%%%%%%%%%%%%%%%%%%%%%%%%%%%%%%

We start our analysis by imposing the flavour symmetry $G_l=U_{cTBM}^\dagger \text{diag}(e^{i\alpha},e^{i\alpha},e^{i\beta}) U_{cTBM}$ into the perturbation term $\delta M^2$. Note that this implies $\delta_1-\delta_4 = \delta_2-\delta_5 = \alpha$, while $\beta = \delta_3-\delta_6 \neq \alpha$. Instead of six CP parameters, going to the $\alpha \alpha \beta$ case restricts the situation to five CP parameters. It is clear that the eigenvector associated to the eigenvalue $e^{i\beta}$ of $G_l$ will also be an eigenvector of $M^2$. However, two independent eigenvectors of $G_l$ with the degenerate eigenvalue $e^{i\alpha}$ will span a subspace of eigenvectors of $G_l$ of which only a particular combination is also eigenvector of $M^2$. Since $G_l$ is diagonalized by $U_{cTBM}^\dagger$, then, after getting rid of the unphysical phases via redefinition of the phases of the charged lepton fields we get
\begin{eqnarray}
 U_{cTBM} (M^2_{cTBM} + \delta M^2) U_{cTBM}^\dagger = \left(
 \begin{array}{ccc}
{\cal M}^2_{11}&\delta{\cal M}^2e^{i \phi}&0\\
\delta{\cal M}^2e^{-i \phi}&{\cal M}^2_{22}&0\\
0 & 0 & m_\tau ^2
\end{array}\right)\,,
\label{eq:Ml12}
\end{eqnarray}
where ${\cal M}^2_{11}$, ${\cal M}^2_{22}$, $\delta{\cal M}^2$ and $\phi$ are real parameters. Notice that the form of the mass matrix is not dependent on the particular values of $\alpha$ and $\beta$, only on the choice of the degenerate (12) sector in this case. Then the imposition of $X_1$ implies $\phi = \frac{\delta_1-\delta_2}{2}$ with $\delta_3$ being completely unphysical. Regarding the CP labels, note that only the combination $\delta_1-\delta_2$ is physical, but not the two phases $\delta_1$ and $\delta_2$ independently. As a consequence, the generalized CP symmetries enforce the charged lepton mass matrix with perturbation to be of the following form,
\begin{eqnarray}
U_{cTBM} (M^2_{cTBM} + \delta M^2) U_{cTBM}^\dagger = \left(
\begin{array}{ccc}
{\cal M}^2_{11}&\delta{\cal M}^2\,e^{i \frac{\delta_1-\delta_2}{2}}&0\\
\delta{\cal M}^2\,e^{-i\frac{\delta_1-\delta_2}{2}}&{\cal M}^2_{22}&0\\
0 & 0 & m_\tau ^2
\end{array}\right)\,.
\end{eqnarray}
As a consistency check, we can see that imposing $X_2$ does not add any new information, since we have already imposed the flavour symmetry parametrized by $\alpha = \delta_1-\delta_4 = \delta_2-\delta_5$ and $\beta= \delta_3-\delta_6$. We can now see that the mass matrix in Eq.~\eqref{eq:Ml12} can be diagonalized by $\text{diag}(e^{i\frac{\delta_1}{2}},e^{i\frac{\delta_2}{2}},e^{i\frac{\delta_3}{2}})U^T_{12}(\theta, 0)$ with
\begin{eqnarray}
\nonumber &&\qquad\qquad\qquad\tan2\theta=-\frac{2\delta{\cal M}^2}{{\cal M}^2_{22}-{\cal M}^2_{11}}\,  , \,\qquad  \delta{\cal M}^2 = -\frac{1}{2} (m_\mu^2-m_e^2) \sin 2\theta \,, \\
&&{\cal M}^2_{11} = \frac{1}{2}[m_e^2(1+\cos 2\theta) + m_\mu^2(1-\cos 2\theta)]\, , ~~\, {\cal M}^2_{22} = \frac{1}{2} [m_e^2(1-\cos 2\theta) + m_\mu^2(1+\cos 2\theta)]\,.
\end{eqnarray}
Note that $\theta$ will always be in the first or fourth quadrant and is expected to be small, since we are perturbing the mass matrix.
Therefore the charged lepton diagonalization matrix is given by
\begin{equation}
U_{cl}=U_{cTBM}^\dagger~ \text{diag}(e^{i\frac{\delta_1}{2}},e^{i\frac{\delta_2}{2}},e^{i\frac{\delta_3}{2}})~ U^T_{12}(\theta, 0)\,.
\end{equation}
Consequently the lepton mixing matrix $U_{lep} = U_{cl}^\dagger U_{\nu} = U_{cl}^\dagger$ will be given by
\begin{equation}
U_{lep}=U_{12}(\theta, 0) ~ \text{diag}(e^{-i\frac{\delta_1}{2}},e^{-i\frac{\delta_2}{2}},e^{-i\frac{\delta_3}{2}})~ U_{cTBM}\,.
\end{equation}
Moreover, we can exploit the relation
\begin{eqnarray}
U_{12}(\theta, 0)~ \text{diag}(e^{-i\frac{\delta_1}{2}},e^{-i\frac{\delta_2}{2}},e^{-i\frac{\delta_3}{2}}) =\text{diag}(e^{-i\frac{\delta_1}{2}},e^{-i\frac{\delta_2}{2}},e^{-i\frac{\delta_3}{2}}) U_{12}\left(\theta,\delta\right)\,,
\end{eqnarray}
with $ \delta=(\delta_2-\delta_1)/2$. Since the phases on the left side of a mixing matrix can be absorbed by redefinitions of the charged lepton fields, only the combination of CP labels given by $\delta\equiv(\delta_2-\delta_1)/2$ will be physically meaningful. The lepton mixing matrix can be written as
\begin{equation}
\boxed{U_{lep}=U_{12}\left(\theta, \delta\right)~ U_{cTBM}}\,.
\label{eq:Ulepaab}
\end{equation}
where we remind that $\theta$ is a free real parameter and $\delta$ is the meaningful CP label. Now we can proceed to extract the mixing parameters from the lepton mixing matrix in Eq.~\eqref{eq:Ulepaab}. In the simplifying limit $\rho \to 0$ and $\sigma \to 0$, i.e. when the CP symmetries of the real TBM matrix $U_{rTBM}^\dagger$ and not $U_{cTBM}^\dagger$ are imposed, the mixing parameters are given by the following expressions
\begin{eqnarray}
\nonumber\sin^2 \theta_{13} & = & \frac{\sin^2\theta}{2}\,,\qquad
\sin^2 \theta_{12}=\frac{2+2\sin2\theta\cos\delta}{3\cos^2\theta+3}\,,\qquad
\sin^2 \theta_{23}=\frac{\cos^2\theta}{\cos^2\theta+1}\,,\\
\nonumber \sin\delta_{CP} & = & \frac{{\rm sign}(\sin2\theta)\left(2\cos^2\theta+2\right)\sin\delta}{
\sqrt{\big(5+3\cos2\theta-4\sin 2\theta\cos\delta\big)
\big(2+2\sin2\theta\cos\delta\big)}}\,,\\
\nonumber \cos\delta_{CP} & = &
\frac{{\rm sign}(\sin2\theta)\left(\sin 2\theta+\left(6\cos^2\theta-2\right)\cos\delta\right)}{
\sqrt{\big(5+3\cos2\theta-4\sin 2\theta\cos\delta\big)
\big(2+2\sin2\theta\cos\delta\big)}}\,,\\
\nonumber \tan \delta_{CP} & = & \frac{(2\cos^2\theta+2)\sin\delta}{\sin2\theta+(6\cos^2\theta-2)\cos\delta}\,,\\
\nonumber \sin 2 \phi_{12} & = & \frac{3\sin\theta(5\cos\theta+3 \cos3\theta)\sin\delta+6\sin^2\theta\cos^2\theta\sin 2\delta}{\left(5+3\cos2\theta-4\sin2\theta\cos\delta\right) \left(1+\sin2\theta\cos\delta\right)}\,,\\
\nonumber \cos 2\phi_{12} & = & 1-\frac{9\sin^22\theta\sin^2\delta}{
(5+3\cos2\theta-4\sin2\theta\cos\delta)(1+\sin2\theta\cos\delta)}\,,\\
\nonumber \sin 2\phi_{13} & = & \frac{8\sin 2\delta \cos^2\theta-4\sin\delta\sin 2\theta}{5+3\cos2\theta-4\sin2\theta\cos\delta}\,,\\
\label{eq:anglesaab}\cos 2\phi_{13} & = & 1-\frac{16\cos^2\theta\sin^2\delta}{5+3\cos2\theta-4\sin2\theta\cos\delta}\,.
\end{eqnarray}
The expressions for the general scenario in which $\rho$ and $\sigma$ are nonzero and the initial symmetry is the complex TBM mixing matrix $U_{cTBM}^\dagger$ are particularly lenghty, specially concerning the CP parameters. However, we can recover the general scenario just by doing the following substitutions in the previous equations
\begin{eqnarray}
&&\delta~\to~\delta-2\rho\,,\quad \phi_{12}~\to~\phi_{12}+\rho\,,\qquad\phi_{13}~\to~\phi_{13}+\rho+\sigma\,.
\label{eq:transaab}
\end{eqnarray}
As a result, the modified variant of the complex TBM matrix predicts the same relations between mixing angles and the Dirac CP phase as the modified variant of the real TBM case. These predictions can be neatly summarized as
\begin{eqnarray}
\label{eq:eq:sumR_clep_1}
\boxed{\cos^2 \theta_{23} \cos^2 \theta_{13}=1/2}\,,\qquad\quad
\boxed{\cos\delta_{CP}=\frac{3\cos2\theta_{12}(2-3\cos^2\theta_{13})+\cos^2\theta_{13}}{6\sin2\theta_{12}\sin\theta_{13}\sqrt{2\cos^2\theta_{13}-1}}}\,.
\end{eqnarray}
This is because, if we only look at the oscillation observables, all the CP parameters ($\rho$, $\sigma$, $\delta_1$, $\delta_2$ and $\delta_3$) are either unphysical for oscillations ($\delta_3$ and $\sigma$), or can be rewritten in terms of just one meaningul CP label, $\delta' \equiv (\delta_2-\delta_1)/2-2\rho$, while the expressions of the oscillation parameters ($\theta_{ij}$ and $\delta_{CP}$) will take the same form as in Eq.~\eqref{eq:anglesaab} just replacing $\delta$ by $\delta'$.

\begin{figure}[t!]
\begin{center}
\begin{tabular}{c}
\includegraphics[width=0.8\linewidth]{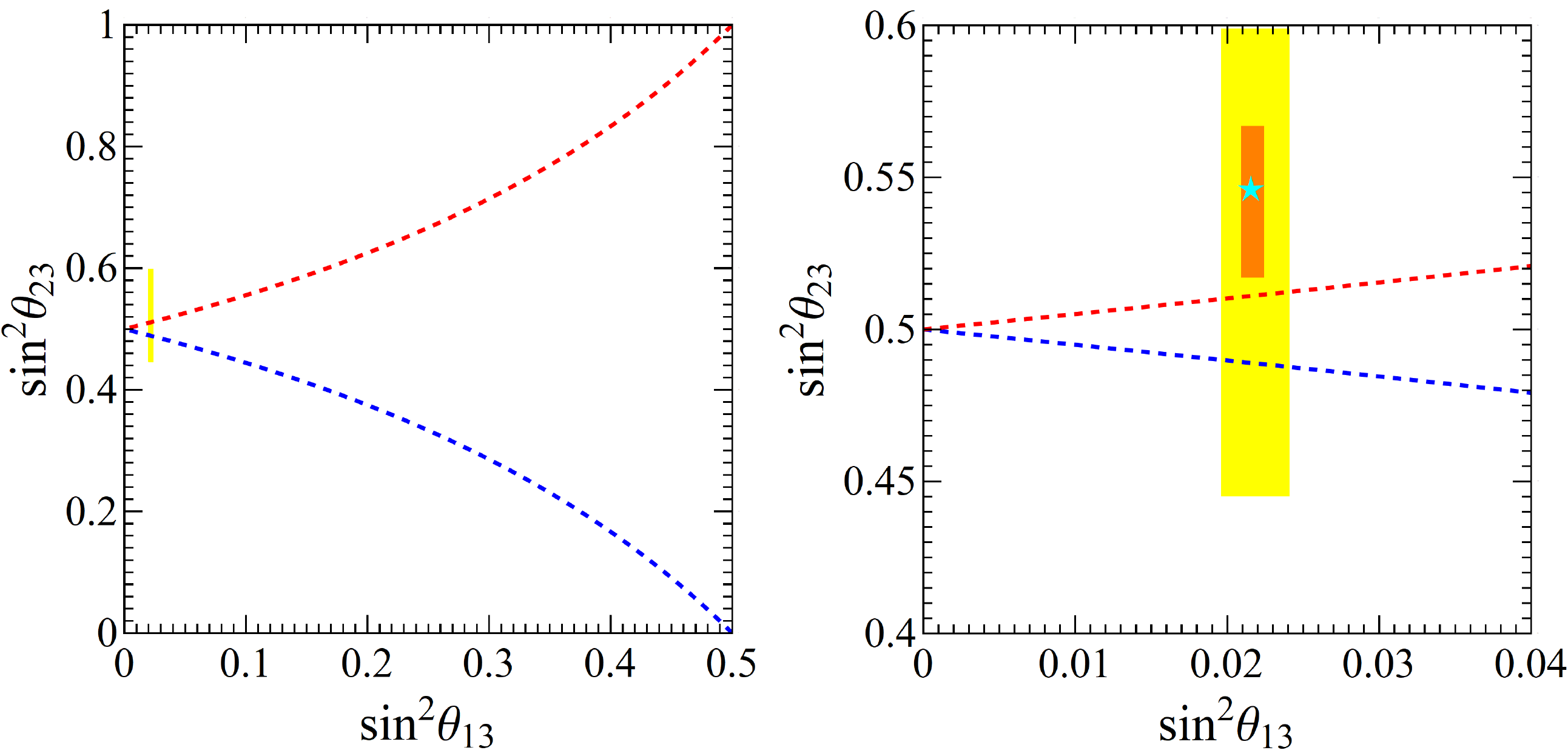}
\end{tabular}
\caption{Correlations between $\theta_{13}$ and $\theta_{23}$ when two CP symmetries are preserved by the charged lepton sector. For the $\alpha\alpha \beta$ case (blue dashed line), they come from Eq.~\ref{eq:eq:sumR_clep_1} (left) and the $\alpha \beta \alpha$ case (red dashed line), they come from Eq.~\ref{eq:eq:sumR_clep_2} (left). The boxes represent the $3\sigma$ and $1\sigma$ allowed ranges for normal ordered neutrino masses~\cite{deSalas:2017kay}. The star is the best-fit value of $\sin^2\theta_{13}$ and $\sin^2\theta_{23}$. Note that fixing $\theta_{13}$ in its $3\sigma$ allowed range greatly restricts the allowed values of the atmospheric angle $\theta_{23}$, nearly maximal in both cases. The $\alpha\alpha \beta$ and $\alpha\beta\alpha $ cases correspond to first and second octant, respectively.}
\label{fig:plot13vs23}
\end{center}
\end{figure}
From Eq.~\eqref{eq:anglesaab} we see that both $\theta_{13}$ and $\theta_{23}$ only depend on the free parameter $\theta$. Hence the possible values of the free parameter $\theta$ are strongly constrained due to a very good measurement of the reactor angle $\theta_{13}$. This implies a very sharp prediction for the atmospheric angle $\theta_{23}$ independent of the particular values of the CP labels $\delta_1$ and $\delta_2$.
The good determination of $\theta_{13}$~\cite{deSalas:2017kay, deSalas:2018bym, Gariazzo:2018pei, globalfit} and the fact that the atmospheric mixing angle $\theta_{23}$ is a slowly varying function of $\theta_{13}$ in the allowed $3\sigma$ range, leads to a tight prediction, $\theta_{23} \in [44.28^{\circ}, 44.43^{\circ}]$, as shown in Fig.~\ref{fig:plot13vs23}.

\begin{figure}[h!]
\begin{center}
\begin{tabular}{cc}
\includegraphics[width=0.8\linewidth]{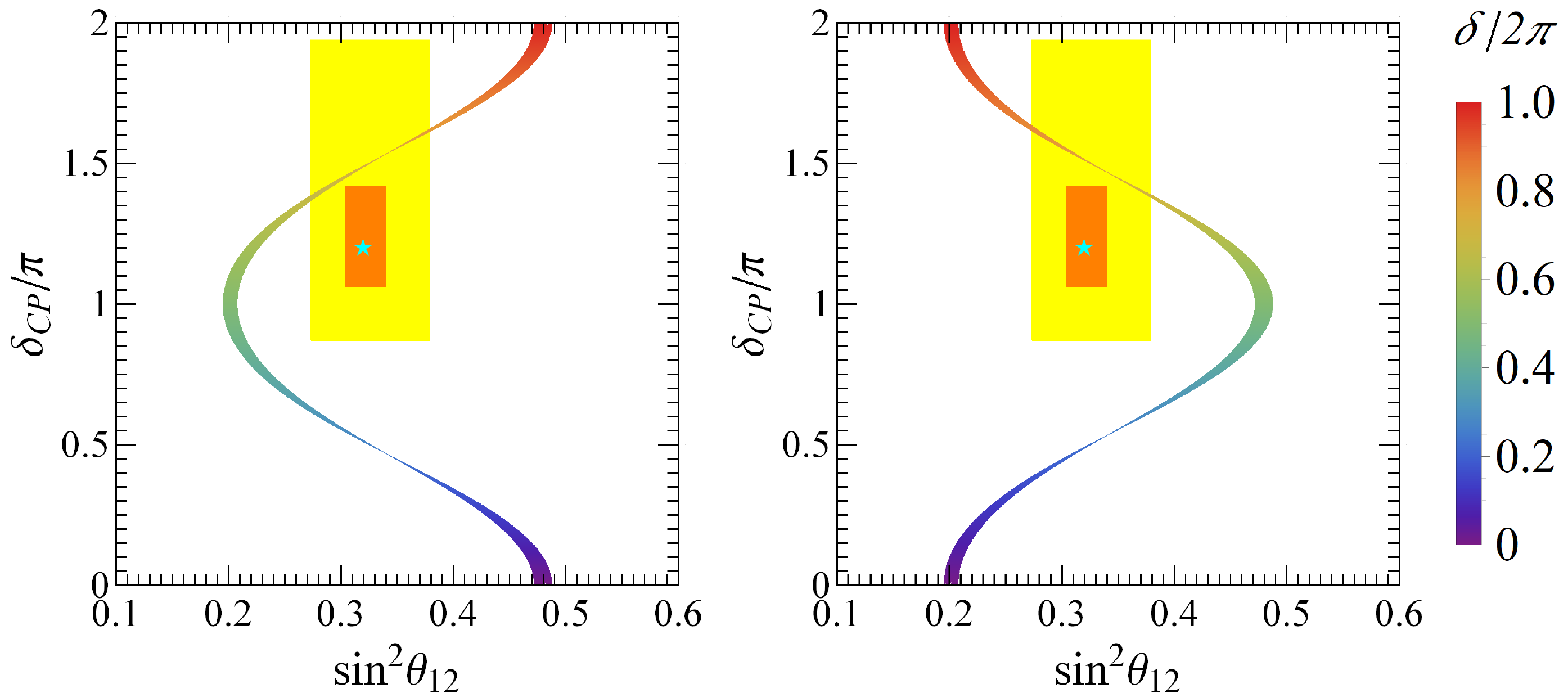}
\end{tabular}
\caption{Correlations between the mixing angle $\theta_{12}$ and the CP phase $\delta_{CP}$ when two CP symmetries are preserved by the charged lepton sector. For the $\alpha \alpha \beta$ case (left panel), they come from Eq.~\ref{eq:eq:sumR_clep_1} (right), and in the $\alpha \beta \alpha$ case (right panel), they come from Eq.~\ref{eq:eq:sumR_clep_2} (right). The boxes are the $3\sigma$ and $1\sigma$ allowed ranges respectively for normal ordered neutrino masses~\cite{deSalas:2017kay}. The star is the best-fit value of $\sin^2\theta_{12}$ and $\delta_{CP}$. The reactor angle $\theta_{13}$ is assumed to lie in the $3\sigma$ region of the global fit~\cite{deSalas:2017kay}.The shaded colour indicates the value of the CP label $\delta$. }
\label{fig:12vsdeltaCP}
\end{center}
\end{figure}

\begin{figure}[h!]
\begin{center}
\begin{tabular}{c}
\includegraphics[width=0.8\linewidth]{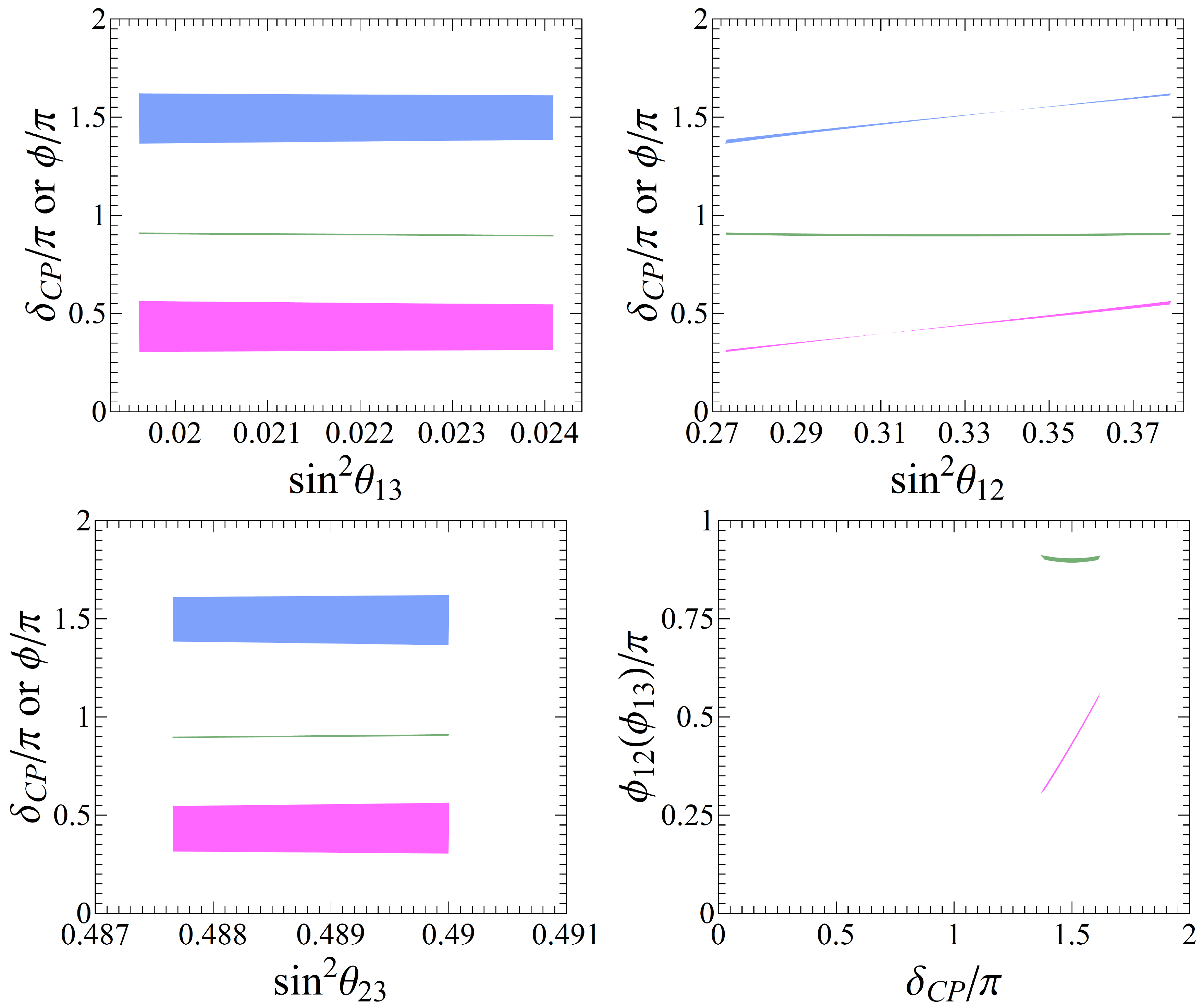}
\end{tabular}
\caption{Correlations between mixing angles and CP phases when two CP symmetries are preserved by the charged lepton sector in the case $\alpha \alpha \beta$. We treat both parameters $\theta$ and $\delta$ as random numbers, the three lepton mixing angles and Dirac CP phase $\delta_{CP}$ are required to lie within their $3\sigma$ ranges~\cite{deSalas:2017kay}. The blue, green and magenta regions correspond to the CP violation phases $\delta_{CP}$, $\phi_{12}$ and $\phi_{13}$, respectively.}
\label{plots123}
\end{center}
\end{figure}

Turning to the solar angle  $\theta_{12}$ Eq.~\eqref{eq:anglesaab} indicates that $\theta_{12}$ and the Dirac CP phase $\delta_{CP}$ depend on both the free parameter $\theta$ and the CP parameter $\delta$. The correlation between $\sin^2\theta_{12}$ and $\delta_{CP}$ is displayed in Fig.~\ref{fig:12vsdeltaCP}, where the values of the CP label $\delta_{CP}$ are indicated by the color shadings. Requiring the solar angle $\theta_{12}$ in the experimentally preferred $3\sigma$ range~\cite{deSalas:2017kay}, we can read out the allowed region of $\delta_{CP}$ is $1.37<\delta_{CP}/\pi<1.62$. Furthermore, we perform a numerical analysis and randomly scan over $\theta$ and the CP label $\delta$ in the ranges $[-\pi/2, \pi/2]$ and $[0, 2\pi]$ respectively, keeping only the points for which the lepton mixing angles $\theta_{ij}$ and $\delta_{CP}$ are consistent with experimental data at $3\sigma$ level. The resulting predictions for the lepton mixing angles and CP violation phases are displayed in Fig.~\ref{plots123}. We observe strong correlations among the solar angle $\sin^2\theta_{12}$, the Dirac CP phase $\delta_{CP}$ and the Majorana phase $\phi_{12}$ and $\phi_{13}$ in this case.

The situation changes for the Majorana phases, since non-zero $\rho$ and $\sigma$ will shift the Majorana phases, as can be seen in Eq.~\eqref{eq:transaab}. Therefore, the difference between the symmetries of the real and the complex versions of TBM can be seen only in neutrinoless double beta decay experiments, as discussed in 
Sec.~\ref{sec:neutr-double-decay}.

%%%%%%%%%%%%%%%%%%%%%%%%%%%%%%%%%%%%%%%%%%%%%%%%%%%%%%%%%%%%%%%%%%%%%%%%%%%%%%%%%%%%%%%%%%%%%%%%%%%%%%%%%%%%%%%%%%%%%%%%%%%%%%%%
\subsubsection{Case $\alpha \beta \alpha$}
\label{sec:p132-p231}
%%%%%%%%%%%%%%%%%%%%%%%%%%%%%%%%%%%%%%%%%%%%%%%%%%%%%%%%%%%%%%%%%%%%%%%%%%%%%%%%%%%%%%%%%%%%%%%%%%%%%%%%%%%%%%%%%%%%%%%%%%%%%%%%%

Following an analogous argument as in the $\alpha \alpha \beta$ case, we can now study the case in which the flavour symmetry is $G_l=U_{cTBM}^\dagger \text{diag}(e^{i\alpha},e^{i \beta},e^{i\alpha}) U_{cTBM}$. In this case the perturbation $\delta M ^2$ satisfies
\begin{equation}
\label{eq:Ml13}
U_{cTBM}(M^2_{cTBM} + \delta M^2) U^\dagger_{cTBM}=
\left(\begin{array}{ccc}
{\cal M}^2_{11}&0&\delta{\cal M}^2\,e^{i \left(\frac{\delta_1-\delta_3}{2} \right)}\\
0 & m_\mu^2 & 0\\
\delta{\cal M}^2\, e^{-i\left(\frac{\delta_1-\delta_3}{2} \right)}&0&{\cal M}^2_{33}
\end{array}\right)\,,
\end{equation}
where ${\cal M}^2_{11}$, ${\cal M}^2_{22}$, $\delta{\cal M}^2$ and ${\cal M}^2_{33}$ are real parameters and $m_\mu$ is the muon mass.
Eq.~\eqref{eq:Ml13} can be diagonalized by $\text{diag}(e^{i\delta_1/2}, e^{i\delta_2/2}, e^{i\delta_3/2})~ U_{13}(\theta, 0)^T$, with
\begin{eqnarray}
\nonumber \tan2\theta=-\frac{2\delta{\cal M}^2}{{\cal M}^2_{22}-{\cal M}^2_{11}}\,  ,& & \, \delta{\cal M}^2 = -\frac{1}{2}(m_\tau^2-m_e^2) \sin 2\theta \\
{\cal M}^2_{11} = \frac{1}{2} [m_e^2(1+\cos 2\theta) + m_\tau^2(1-\cos 2\theta)]\, , & & \, {\cal M}^2_{33} = \frac{1}{2} [m_e^2(1-\cos 2\theta) + m_\tau^2(1+\cos 2\theta)]\,,
\end{eqnarray}
where $\theta$ is expected to be small. Using similar arguments to those of the previous case, the lepton mixing matrix $U_{lep} = U_{cl}^\dagger U_\nu$ is given by
\begin{equation}
\boxed{U_{lep}=U_{13}\left(\theta,\delta\right) ~  U_{cTBM}}\,,
\label{eq:Ulepaba}
\end{equation}
where we have defined $\delta \equiv (\delta_3-\delta_1)/2$. Extracting the mixing parameters in the real TBM case where $\rho \to 0$ and $\sigma \to 0$, we find them to be
\begin{eqnarray}
\nonumber \sin^2 \theta_{13}  & = &  \frac{\sin^2\theta}{2}\,,\qquad
\sin^2 \theta_{12} = \frac{2-2\sin 2\theta\cos\delta}{3\cos^2\theta+3}\,,\qquad
\sin^2 \theta_{23} = \frac{1}{\cos^2\theta+1}\,,\\
\end{eqnarray}
while the CP phases are given by
\begin{eqnarray}
\nonumber \sin\delta_{CP} & = &
\frac{{\rm sign}(\sin2\theta)\left(2\cos^2\theta+2\right)\sin\delta}{
\sqrt{\big(5+3\cos2\theta+4\sin 2\theta\cos\delta\big)
\big(2-2\sin2\theta\cos\delta\big)}}\,,\\
\nonumber \cos\delta_{CP} & = &-
\frac{{\rm sign}(\sin2\theta)\big(\sin 2\theta-\left(6\cos^2\theta-2\right)\cos\delta\big)}{
\sqrt{\big(5+3\cos2\theta+4\sin 2\theta\cos\delta\big)
\big(2-2\sin2\theta\cos\delta\big)}}\,,\\
\nonumber \tan \delta_{CP}  & = &  \frac{-2(\cos^2\theta+1)\sin\delta}
{\sin 2\theta-2(3\cos^2\theta-1)\cos\delta}\,,\\
\nonumber \sin 2 \phi_{12} & = & \frac{-3\sin\theta(5\cos\theta+3 \cos3\theta)\sin\delta+6\sin^2\theta\cos^2\theta\sin2\delta}{(5+3\cos2\theta+4\sin2\theta\cos\delta) (1-\sin2\theta\cos\delta)}\,,\\
\nonumber\cos 2\phi_{12} & = & 1-\frac{9\sin^22\theta\sin^2\delta}{
(5+3\cos2\theta+4\sin2\theta\cos\delta)(1-\sin2\theta\cos\delta)}\,,\\
\nonumber \sin 2\phi_{13} & = & \frac{8\sin\delta\cos^2\theta+4\sin\delta\sin 2\theta}{5+3\cos2\theta+4\sin2\theta\cos\delta}\,,\\
\label{eq:anglesaba} \cos 2\phi_{13} & = & 1-\frac{16\cos^2\theta\sin^2\delta}{5+3\cos2\theta+4\sin2\theta\cos\delta}\,.
\end{eqnarray}
Similar to the previous case, the general scenario in which $\rho$ and $\sigma$ are nonzero can be recovered by making the following substitutions
\begin{eqnarray}
&&\delta~\to~\delta-2\rho-2\sigma\,,\quad \phi_{12}~\to~\phi_{12}+\rho\,,\quad\phi_{13}~\to~\phi_{13}+\rho+\sigma\,.
\label{eq:transaba}
\end{eqnarray}

\begin{figure}[t]
\begin{center}
\begin{tabular}{cc}
\includegraphics[width=0.8\linewidth]{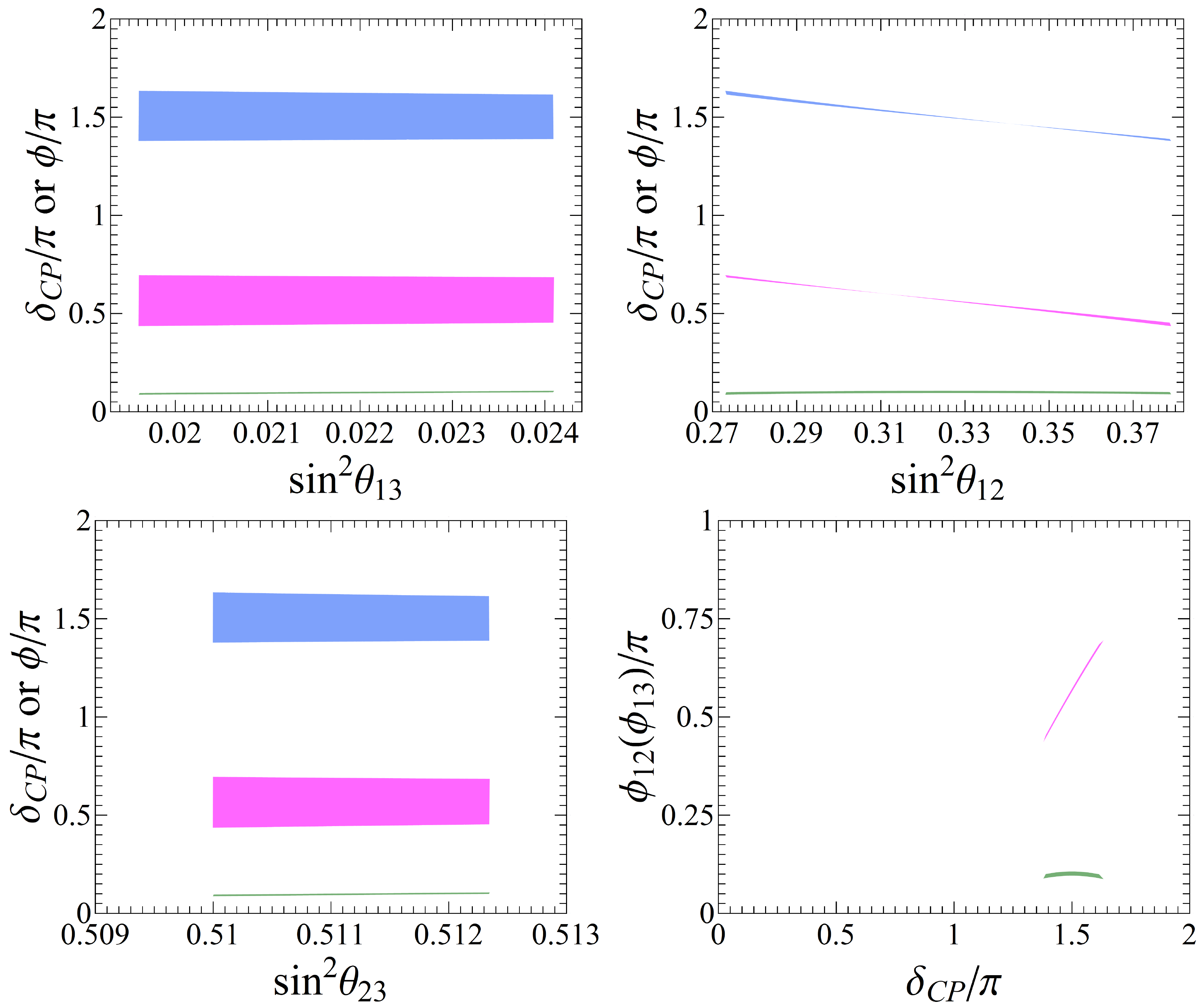}
\end{tabular}
\caption{Correlations between mixing angles and CP phases when two CP symmetries are preserved by the charged lepton sector for the $\alpha \beta\alpha$ case.
The blue, green and magenta regions  correspond to $\delta_{CP}$, $\phi_{12}$ and $\phi_{13}$, respectively.}
\label{plots132}
\end{center}
\end{figure}

Again, as in the previous case, the good measurement of $\theta_{13}$ severely restrict the allowed values of the free parameter $\theta$. This gives a very sharp prediction for the atmospheric angle $\theta_{23}\in[45.56^\circ, 45.71^\circ]$, as can be seen in Fig.~\ref{fig:plot13vs23}. Moreover, we find the following analytical correlations between the physical parameters
\begin{eqnarray}
\label{eq:eq:sumR_clep_2}
\boxed{\sin^2 \theta_{23} \cos^2 \theta_{13}=1/2}\,,\qquad\quad
\boxed{\cos\delta_{CP}=\frac{3\cos2\theta_{12}(3\cos^2\theta_{13}-2)-\cos^2\theta_{13}}{6\sin2\theta_{12}\sin\theta_{13}\sqrt{2\cos^2\theta_{13}-1}}}\,,
\end{eqnarray}
which also hold in the case where $\rho$ and $\sigma$ are non-zero. Notice that, in contrast to the $\alpha \alpha \beta$ case, this time $\theta_{23}$ falls in the second octant, although again very close to the maximal mixing value.
It is worth noting that the two cases show a very similar behaviour, as can be seen from figures~\ref{fig:plot13vs23} and \ref{fig:12vsdeltaCP}. In both cases the angle $\theta_{23}$ is very close to the maximal mixing value.
As seen in Fig.~\ref{fig:plot13vs23} the first octant corresponds to the $\alpha \alpha \beta$ case, while the $\alpha \beta \alpha$ case
is associated to the second octant. The correlation between $\theta_{12}$ and $\delta_{CP}$ is similar to the previous case but of opposite shape. From here we can read off the allowed range for the Dirac CP phase, i.e. $1.379\pi\leq\delta_{CP}\leq1.634\pi$.
The numerical results for the lepton mixing and CP phase parameters for the real TBM case are shown in Fig.~\ref{plots132}, where the three mixing angles and the Dirac CP phase are required to lie inside the current $3\sigma$ range~\cite{deSalas:2017kay}. Similar to the 
$\alpha\alpha\beta$ case, the modified variants of the complex and real TBM obey the same relations given in Eq.~\eqref{eq:eq:sumR_clep_2} and hence their predictions for the three mixing angles and the Dirac CP phase are the same. However they differ in their predictions for neutrinoless double beta decay, as discussed in Sec. \ref{sec:neutr-double-decay}.

%%%%%%%%%%%%%%%%%%%%%%%%%%%%%%%%%%%%%%%%%%%%%%%%%%%%%%%%%%%%%%%%%%%%%%%%%%%%%%%%%%%%%%%%%%%%%%%%%%%%%%%%%%%%%%%%%%%%%%%%%
\section{Charged lepton mass matrix conserving one CP symmetry}
\label{sec:one-cl}
%%%%%%%%%%%%%%%%%%%%%%%%%%%%%%%%%%%%%%%%%%%%%%%%%%%%%%%%%%%%%%%%%%%%%%%%%%%%%%%%%%%%%%%%%%%%%%%%%%%%%%%%%%%%%%%%%%%%%%%%%%

We will now study the case in which only one CP symmetry is preserved in the perturbation term. The CP symmetries compatible
with $U_{cTBM}$ in the charged sector are
\begin{equation}
X=U_{cTBM}^\dagger\text{diag}(e^{i\delta_1},e^{i\delta_2},e^{i\delta_3}) U_{cTBM}^\ast
\end{equation}
where $\delta_i$ are CP parameters labelling the CP transformation.
The charged lepton squared mass matrix satisfies the relation
\begin{eqnarray}
X^\dagger M^2 X=M^{2 \ast}\,,
\end{eqnarray}
where $M^2 = M^2_{cTBM}+ \delta M^2$. Then, the matrix form of $U_{cTBM} M^2 U_{cTBM}^\dagger$ can be written as
\begin{equation}
\label{eq:oct9}
U_{cTBM}(M^2_{cTBM}+ \delta M^2) U_{cTBM}^\dagger=
\left(\begin{array}{ccc}
{\cal M}^2_{11}&\delta{\cal M}^2_{12}e^{\frac{i(\delta_1-\delta_2)}{2}}&\delta{\cal M}^2_{13}e^{\frac{i(\delta_1-\delta_3)}{2}}\\
\delta{\cal M}^2_{12}e^{-\frac{i(\delta_1-\delta_2)}{2}}&{\cal M}^2_{22}&\delta{\cal M}^2_{23}e^{\frac{i(\delta_2-\delta_3)}{2}}\\
\delta{\cal M}^2_{13}e^{-\frac{i(\delta_1-\delta_3)}{2}}&\delta{\cal M}^2_{23}e^{-\frac{i(\delta_2-\delta_3)}{2}}&{\cal M}^2_{33}
\end{array}\right)\,,
\end{equation}
where ${\cal M}^2_{11}$, ${\cal M}^2_{22}$, ${\cal M}^2_{33}$ and $\delta{\cal M}^2_{ij}$ are real parameters,
unconstrained by the residual symmetry. The matrix $U_{cTBM} M^2 U_{cTBM}^\dagger$ in Eq.~\eqref{eq:oct9} can be diagonalized by $\text{diag}(e^{i\frac{\delta_1}{2}},e^{i\frac{\delta_2}{2}},e^{i\frac{\delta_3}{2}})~ O_{3\times 3}^T$. Here $O_{3\times 3}$ is a real orthogonal matrix, its matrix elements are determined by the ${\cal M}^2_{11}$, ${\cal M}^2_{22}$, ${\cal M}^2_{33}$ and $\delta{\cal M}^2_{ij}$~\cite{Chen:2018zbq}, and it can be parametrized as
\begin{eqnarray}
 O_{3\times 3} = U_{23}(\theta_1, 0) ~ U_{13}(\theta_2, 0) ~ U_{12}(\theta_3, 0)\,.
\end{eqnarray}
Then, one can write
\begin{equation}
U_{cl}=U_{cTBM}^\dagger\text{diag}(e^{i\frac{\delta_1}{2}},e^{i\frac{\delta_2}{2}},e^{i\frac{\delta_3}{2}})O_{33}^T\,,
\end{equation}
and therefore the lepton mixing matrix can be written as
\begin{equation}
U_{lep}=U_{cl}^\dagger U_\nu = O_{3\times 3}\text{diag}(e^{-i\frac{\delta_1}{2}},e^{-i\frac{\delta_2}{2}},e^{-i\frac{\delta_3}{2}})~ U_{cTBM}\,.
\end{equation}
Extracting the mixing angles in the simple limit of real TBM, i.e. $\rho = 0 = \sigma$ yields
\begin{eqnarray}
\nonumber\sin^2 \theta_{13} & = &  \frac{1}{2} \Big(1+\sin2\theta_2\sin\theta_3\cos\frac{\delta_2-\delta_3}{2}-\cos^2\theta_2\cos^2\theta_3\Big)\,,\\
\nonumber \sin^2 \theta_{12}  & = &  \frac{2 \left(1-\sin2\theta_2 \left(\cos \theta_3 \cos \frac{\delta_1-\delta_3}{2}+\sin \theta_3 \cos \frac{\delta_2-\delta_3}{2}\right)+\sin 2\theta_3 \cos^2\theta_2 \cos\frac{\delta_1-\delta_2}{2}\right)}{3 \left(1-\sin2\theta_2\sin\theta_3\cos\frac{\delta_2-\delta_3}{2}+\cos^2\theta_2 \cos^2\theta_3\right)}\,,\\
\sin^2 \theta_{23}  & = & \sin^2\theta_1-\frac{-2\cos\frac{\delta_2-\delta_3}{2}\sin 2\theta_1 \cos \theta_2 \cos \theta_3
-2\cos2\theta_1\cos^2\theta_3+\sin2\theta_1\sin\theta_2\sin2\theta_3}{2(1-\sin2\theta_2\sin\theta_3\cos\frac{\delta_2-\delta_3}{2}+\cos^2\theta_2 \cos^2\theta_3)}\,.
\end{eqnarray}
One sees that in this case there is no predictivity for the mixing angles, since all parameters are completely free. Since the expressions for the CP parameters are too lengthy to be enlightening, we will not show them here.

\section{Phenomenological Implications}
\label{sec:phen-impl}

In this section we shall study the phenomenological implications of the above CP symmetries for neutrino oscillation as well as neutrinoless double beta decay experiments. We will focus on the case where two remnant CP symmetries are preserved in the charged lepton sector, as this is the most predictive situation.

\subsection{Neutrino oscillations}
\label{sec:neutr-oscill}

The observation of neutrino oscillations indicates that neutrinos are massive and that neutrino flavor eigenstates mix with each other. The three lepton mixing angles and the neutrino mass-squared differences have been precisely measured. However, we have not yet established with high significance whether CP is violated in the lepton sector. Moreover, we still don't know whether the neutrino mass spectrum has normal ordering (NO) or inverted ordering (IO), nor whether the atmospheric angle lies in the first or second octant. The upcoming reactor and long-baseline experiments such as JUNO, DUNE, T2HK should be able to shed light on these issues and they expect to bring us increased precision on the oscillation parameters $\theta_{12}$, $\theta_{23}$ and $\delta_{CP}$. As shown in previous sections, lepton mixing parameters are predicted to lie in narrow regions and correlations among the mixing parameters are obtained when two residual CP symmetries are preserved.  This would translate into phenomenological implications for the expected neutrino and anti-neutrino appearance probabilities in neutrino oscillation experiments.

\begin{figure}[h!]
\begin{center}
\begin{tabular}{cc}
%\hskip-0.6in
\includegraphics[width=0.9\linewidth]{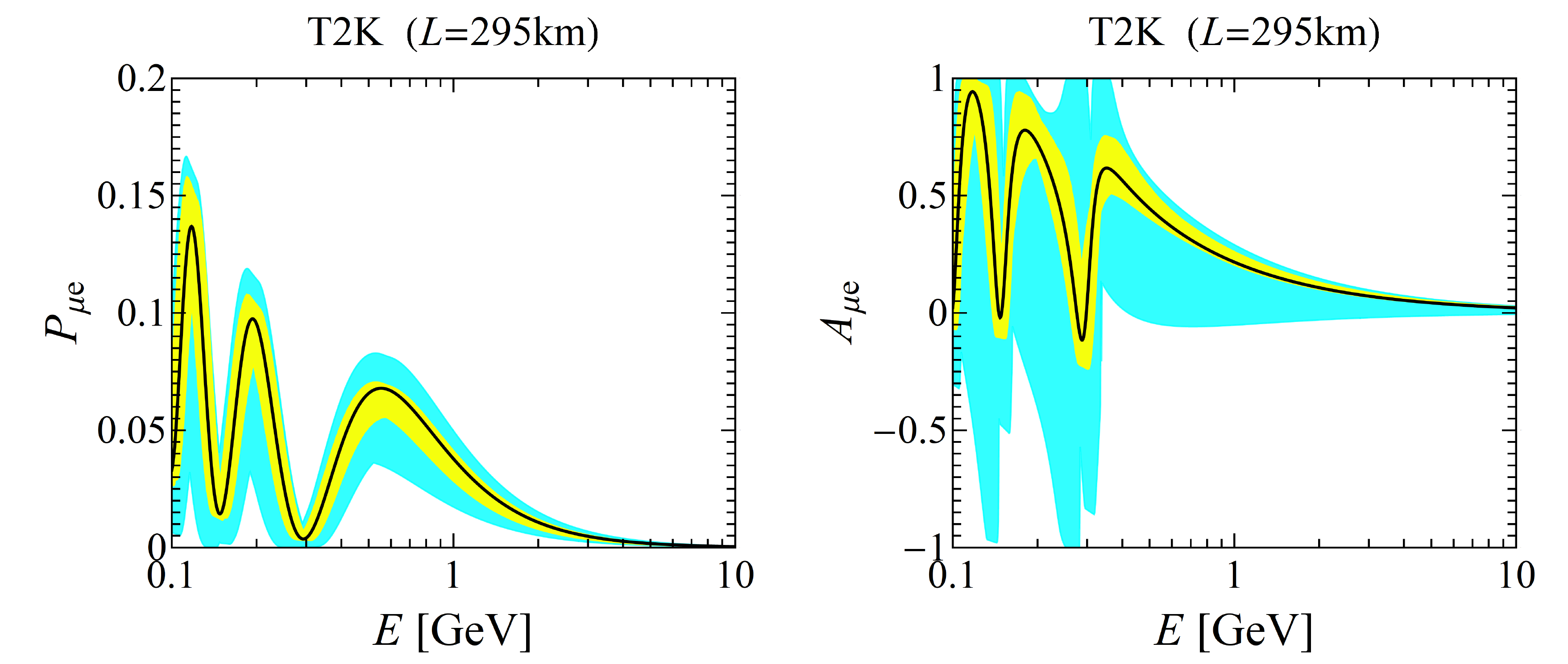}  %&
\end{tabular}
\caption{
The appearance probability $P_{\mu e}$ (left panel) and the CP asymmetry $A_{\mu e}$ (right panel) as functions of the beam energy when the baseline is fixed to 295km (T2K experiment). In both panels the predicted band in cyan is the generically expected region obtained 
by varying the oscillation parameters within their $3\sigma$ ranges~\cite{deSalas:2017kay}, while the yellow band is the prediction for the $\alpha \alpha \beta$ neutrino mixing pattern, see text for explanation.}
\label{fig:oscillationT2k}
\end{center}
\end{figure}

\begin{figure}[h!]
\begin{center}
\begin{tabular}{cc}
%\hskip-0.6in
\includegraphics[width=0.9\linewidth]{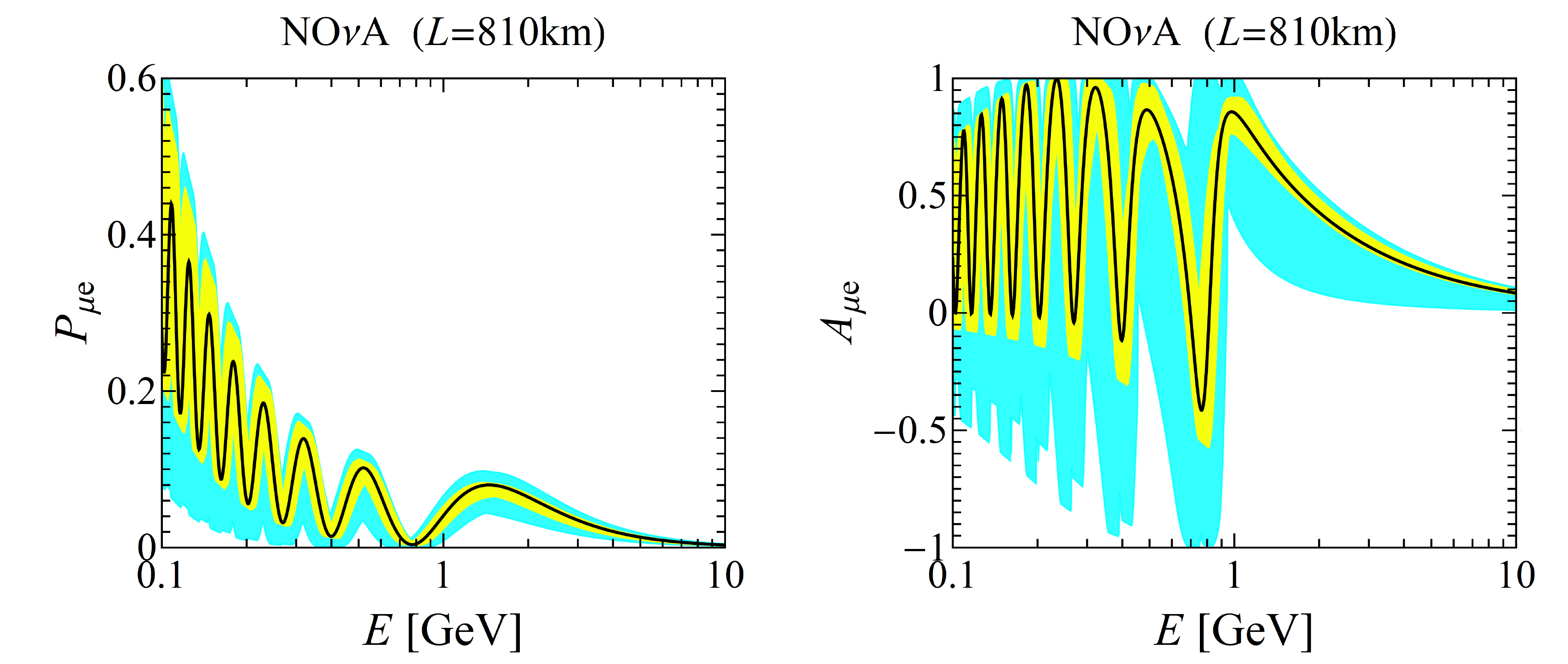}  %&
\end{tabular}
\caption{
{Same as Fig.~\ref{fig:oscillationT2k} when the baseline is fixed to 810km (NOvA experiment), see text for explanation.}
 }
\label{fig:oscillationNova}
\end{center}
\end{figure}

\begin{figure}[h!]
\begin{center}
\begin{tabular}{cc}
%\hskip-0.6in
\includegraphics[width=0.9\linewidth]{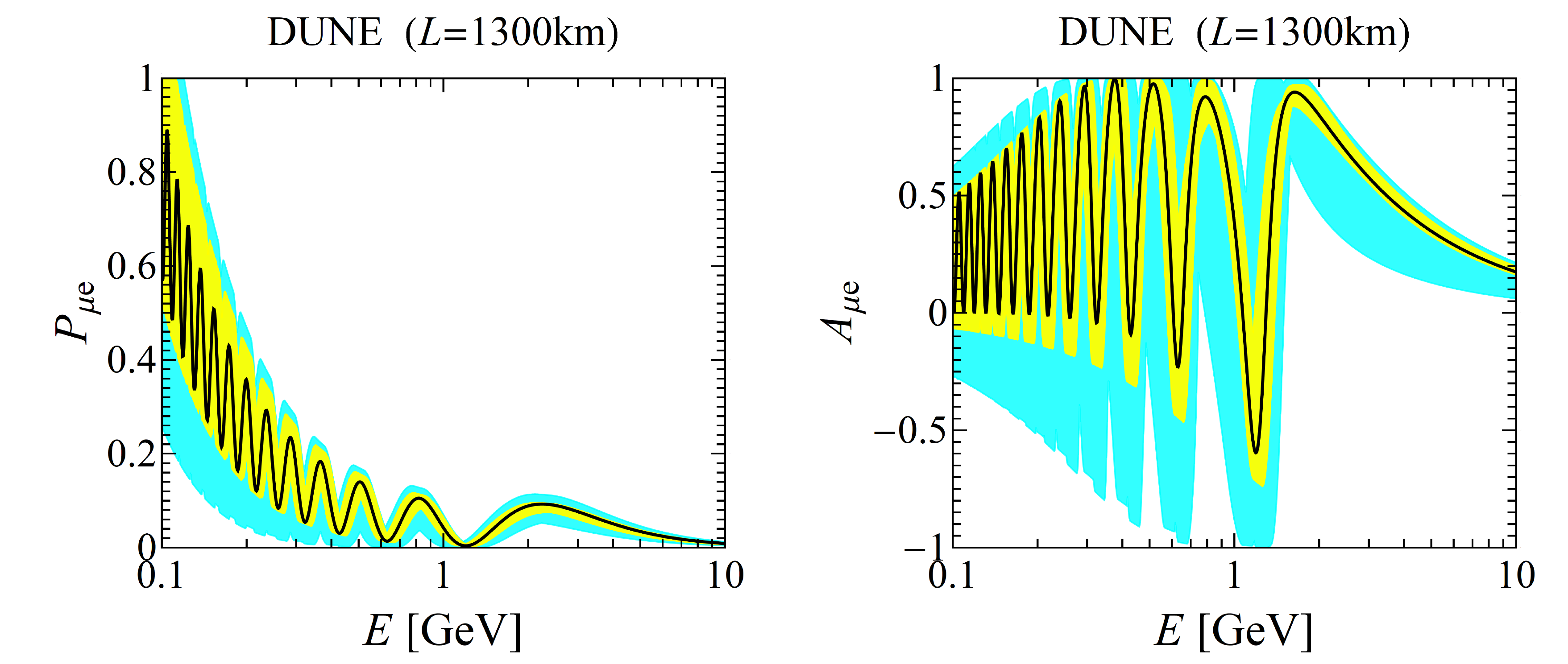}  %&
\end{tabular}
\caption{
{Same as Fig.~\ref{fig:oscillationT2k} when the baseline is fixed to 1300km (DUNE experiment), see text for explanation.}
 }
\label{fig:oscillationDune}
\end{center}
\end{figure}

The $\nu_\mu\to\nu_e$ neutrino oscillation probability in matter can be expanded to second order in the mass hierarchy parameter $\alpha\equiv\Delta m_{21}^2/\Delta m_{31}^2$ and the reactor angle $\sin\theta_{13}$, as follows from~\cite{Akhmedov:2004ny}:
\begin{eqnarray}
\begin{aligned}
P_{\mu e}&=\alpha^2\,\sin^2 2\theta_{12}\,c_{23}^2 \frac{\sin^2A\Delta}{A^2}+4\,s_{13}^2\,s_{23}^2\frac{\sin^2(A-1)\Delta}{(A-1)^2}\\
&\qquad+2\,\alpha\,s_{13}\,\sin2\theta_{12}\,\sin2\theta_{23}\cos(\Delta+\delta_{\rm CP})\, \frac{\sin A\Delta}{A}\,\frac{\sin(A-1)\Delta}{A-1}\,,
\end{aligned}
\end{eqnarray}
with $\Delta=\frac{\Delta m_{31}^2 L}{4 E}$ and $A=\frac{2E V}{\Delta m_{31}^2}$, where $L$ is the baseline length and $E$ is the energy of the neutrino beam.
The matter-induced effective potential is $V\simeq7.56\times10^{-14}\frac{\rho}{\mathrm{g}/\mathrm{cm}^3}\,Y_e\,\mathrm{eV}$ where $Y_e=0.5$ is the electron fraction and a constant matter density $\rho=3\mathrm{g}/\mathrm{cm}^3$ is assumed. The oscillation probability for antineutrinos is related to that for neutrinos by $P_{\bar{\mu}\bar{e}}=P_{\mu e}(\delta_{CP}\to-\delta_{CP},V\to-V)$.
The oscillation probability asymmetry between neutrinos and anti-neutrinos is defined as:
\begin{eqnarray}
A_{\mu e}=\frac{P_{\mu e}-P_{\bar{\mu}\bar{e}}}{P_{\mu e}+P_{\bar{\mu}\bar{e}}}\,.
\end{eqnarray}
In order to illustrate our points we focus, for definiteness, on the revamped $\alpha\alpha\beta$ scenario.
We display the results for the neutrino appearance oscillation probability $P_{\mu e}$ and the CP asymmetry $A_{\mu e}$ in Figs.~\ref{fig:oscillationT2k}, \ref{fig:oscillationNova} and \ref{fig:oscillationDune}. 
These are given in terms of the neutrino energy at a fixed distance of $L=295$ km, $L=810$ km and $L=1300$ km, corresponding to the baselines of the T2K, NOvA and DUNE experiments, respectively. 

Note that our generalized CP symmetries lead to correlations involving the mixing and CP violation parameters. In particular, the three mixing angles and the CP phase can be given in terms of only two parameters. 
This translates into restrictions on the attainable ranges for the neutrino oscillation probabilities. In Figs.~\ref{fig:oscillationT2k}, \ref{fig:oscillationNova} and \ref{fig:oscillationDune} the cyan bands are the generically expected regions obtained when the oscillation parameters are varied within their current $3\sigma$ ranges, while the yellow bands denote the predictions for the $\alpha \alpha \beta$ neutrino mixing pattern. The solid black lines correspond to the current best fit point predictions, where we have chosen the Dirac phase $\delta_{CP}=1.5\pi$. As a final comment let us mention that, by looking at Eqs. \ref{eq:anglesaab} and \ref{eq:transaab} one sees that the oscillation results obtained by taking real or complex TBM as the starting point before revamping, i.e. taking $\rho$ and $\sigma$ in Eq.\ref{eq:ctbm} to be non-zero, are essentially the same.

\subsection{Neutrinoless double decay}
\label{sec:neutr-double-decay}

The neutrinoless double beta ($0\nu\beta\beta$) decay $(A, Z) \to (A, Z + 2) + 2 e^-$ is the unique probe of the Majorana nature of neutrinos~\cite{Schechter:1981bd}.
There are many experiments currently searching for $0\nu\beta\beta$ decay, or in various stages of planning and construction. The sensitivity to this rare process should improve significantly, with good prospects for probing the whole region of parameter space associated with the inverted ordering spectrum. The $0\nu\beta\beta$ decay provides another test of the CP symmetries of TBM. For our $\alpha\alpha\beta$ \textit{Ansatz} in the general scenario where the phases $\rho$ and $\sigma$ are non-zero, we find the analytical expression of the effective Majorana mass $m_{ee}$ is given by
\begin{eqnarray}
m_{ee,\alpha \alpha \beta}^{cTBM} = \frac{1}{6} \left|m_1 e^{2 \,i\rho} \left(\sin \theta-2 e^{i (\delta -\rho)} \cos \theta \right)^2+
2 m_2 \left(e^{i (\delta-\rho) } \cos \theta + \sin \theta \right)^2+3 m_3 e^{-2 i \sigma } \sin^2\theta \right|\,.
\end{eqnarray}
Notice that $\rho$, $\sigma$ and $\delta$ are CP parameters, i.e, they label the CP symmetries respected by the mass matrix, while $\theta$ is a completely free parameter, which represents the degree up to which the mass matrix is not determined by the remnant CP symmetry. If all the three labels $\rho$, $\sigma$ and $\delta$ are treated as free parameters, i.e. scanning over the full class of mass matrices which allow some preserved CP symmetry of the complex TBM \textit{Ansatz}, there is no prediction for the $0\nu\beta\beta$ decay, since the correlations between Majorana phases and mixing angles disappear. It is easy to understand this behaviour from Eq.~\eqref{eq:transaab}.
It is worth noting that in the real TBM case, i.e. $\rho=0=\sigma$, the analytical expression of the effective Majorana mass $m_{ee}$ simplifies to,
\begin{eqnarray}
m_{ee,\alpha \alpha \beta}^{rTBM} = \frac{1}{6} \left|m_1  \left(\sin \theta-2 e^{i \delta} \cos \theta \right)^2+2 m_2
\left(e^{i \delta } \cos \theta + \sin \theta \right)^2+3 m_3  \sin^2\theta \right|\,.
\end{eqnarray}

\begin{figure}[h!]
\begin{center}
\begin{tabular}{cc}
\includegraphics[width=0.45\linewidth]{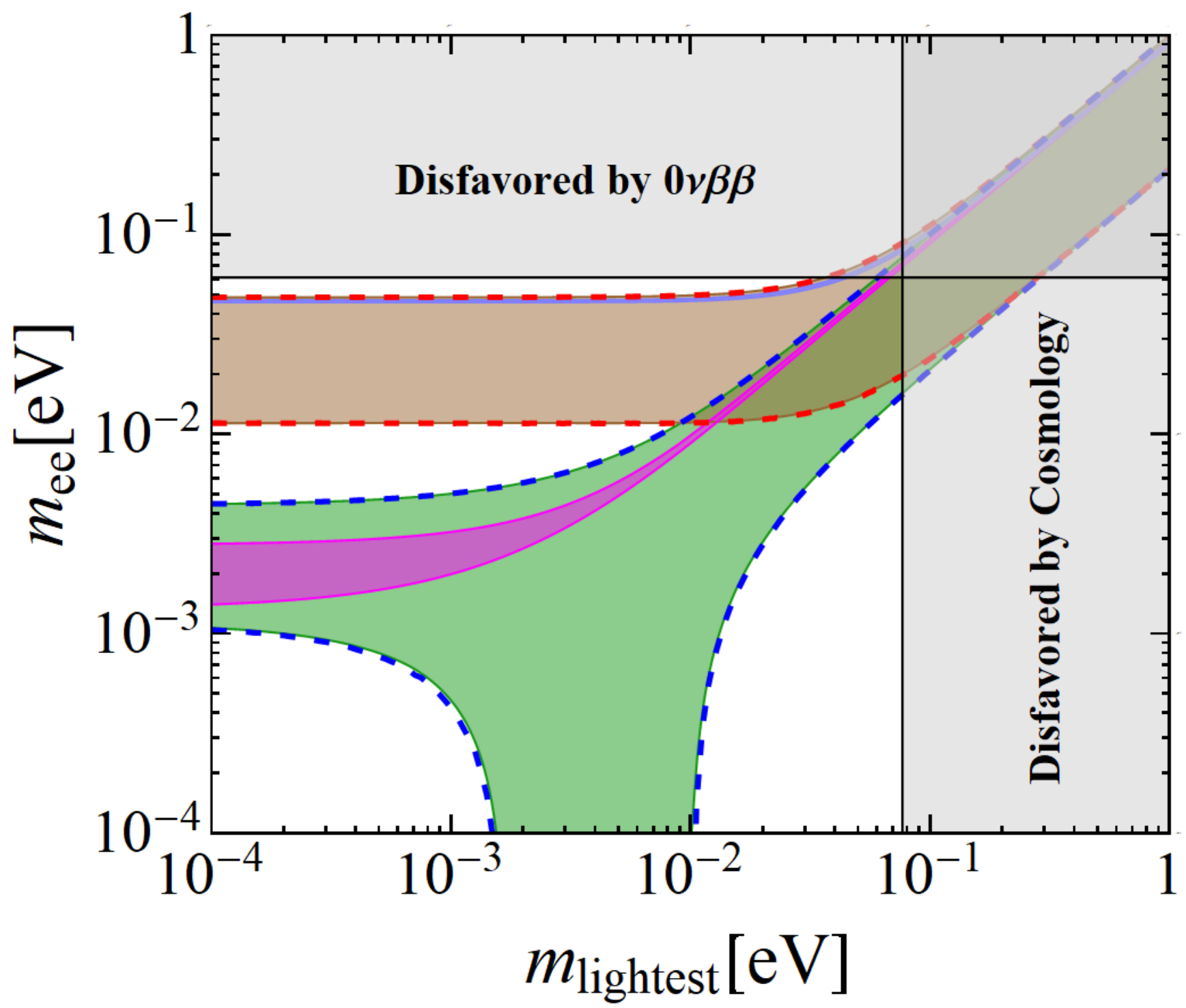}
\includegraphics[width=0.45\linewidth]{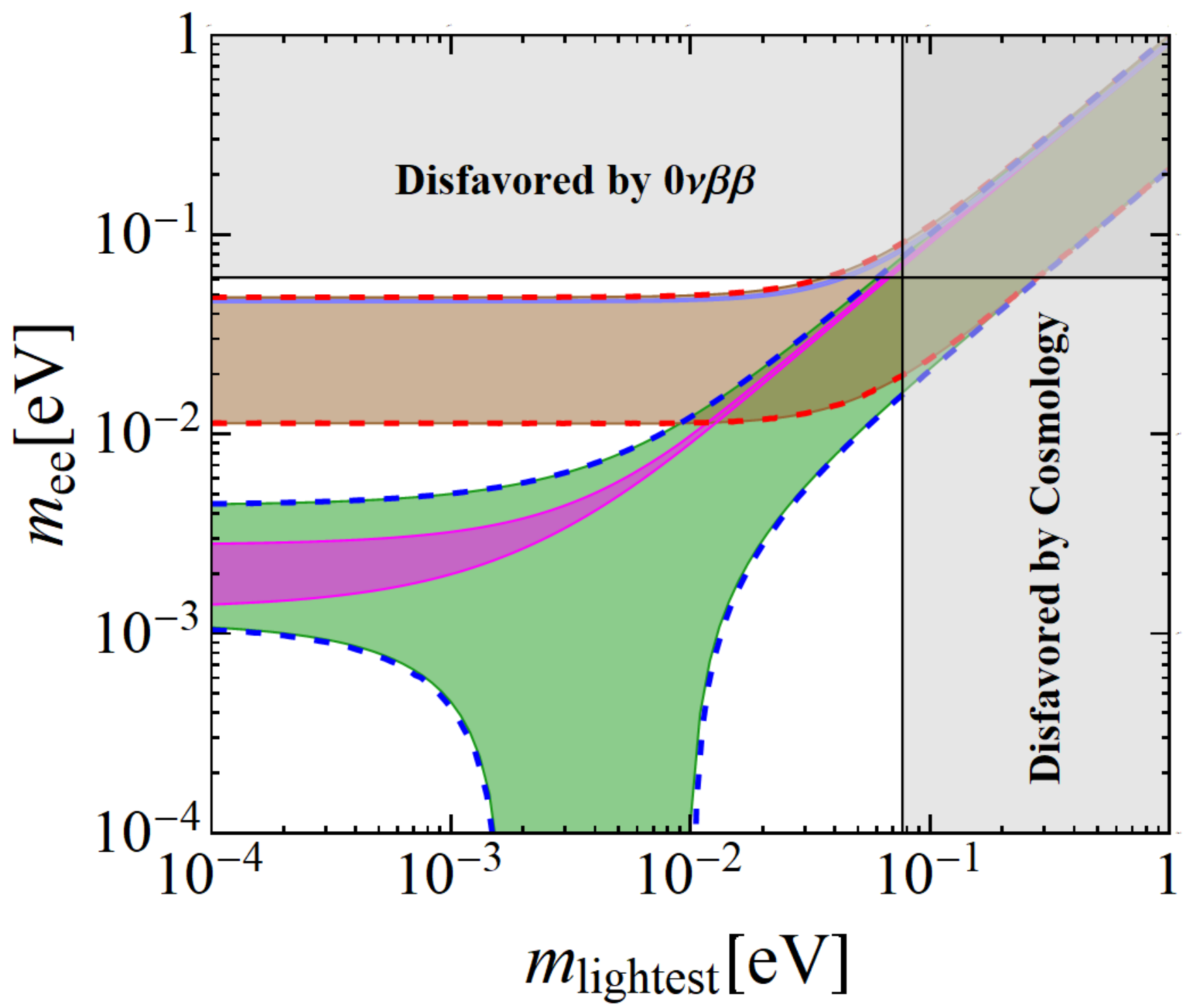}
\end{tabular}
\caption{The effective Majorana mass $m_{ee}$ versus the lightest neutrino mass for the $\alpha\alpha\beta$ (left) and the $\alpha\beta\alpha$ cases (right).
The thick colored regions correspond to the predictions of the complex TBM scenario, while the thin magenta and purple bands are for the real TBM
case, with $\rho=\sigma=0$. The red and blue dashed lines delimit the most general allowed regions for IO and NO neutrino mass spectra, and are obtained by varying
the mixing parameters over their $3\sigma$ ranges~\cite{deSalas:2017kay}. The most stringent current upper limit $m_{ee}<0.061$ eV from
KamLAND-ZEN~\cite{KamLAND-Zen:2016pfg} and EXO-200~\cite{Albert:2017owj} is shown by horizontal grey band. The vertical grey band refers to the current sensitivity of
cosmological data from the Planck collaboration~\cite{Ade:2015xua}.
}
\label{fig:nu0eeaab}
\end{center}
\end{figure}

Limiting ourselves to the real TBM case we can not only generate CP violation as shown in the previous sections, but also obtain very stringent predictions for neutrinoless double beta decay. The allowed values of $m_{ee}$ are shown in Fig.~\ref{fig:nu0eeaab}, where the oscillation parameters are required to lie in the currently preferred $3\sigma$ ranges~\cite{deSalas:2017kay}. One sees that the real TBM prediction for $m_{ee}$ in the IO case corresponds to the upper boundary of the generic IO region, very close to the sensitivities of the upcoming $0\nu\beta\beta$ decay experiments. Notice the existence of a lower bound for $m_{ee}$ in the real TBM scenario also for the NO case, where generically it is absent due to possible cancellations. \\

We can also study the predictions for neutrinoless double beta decay within the $\alpha\beta\alpha$ scenario. The effective mass $m_{ee}$ parameters in this case is given by
% \begin{eqnarray}
% m_{ee, \alpha\beta\alpha}^{cTBM} = \frac{1}{6}\left|m_1 \left(2 \cos \theta e^{i  \sigma} +\sin \theta  e^{-i (\rho + \sigma) }\right)^2+ 2m_2 \left(e^{i (\delta-\rho)}\cos\theta -e^{i\sigma}\sin\theta\right)^2+3 m_3\sin^2\theta \right|\,,
% \end{eqnarray}
\begin{eqnarray}
m_{ee, \alpha\beta\alpha}^{cTBM} = \frac{1}{6}\left|m_1 \left(2 \cos \theta e^{i\delta} +\sin \theta  e^{i (\rho + \sigma) }\right)^2+ 2m_2 \left(e^{i (\delta-\rho)}\cos\theta -e^{i\sigma}\sin\theta\right)^2+3 m_3\sin^2\theta \right|\,,
\end{eqnarray}
which simplifies to the following expression in the limit of $\rho=0=\sigma$,
%
% \begin{eqnarray}
% m_{ee, \alpha\beta\alpha}^{rTBM}=\frac{1}{6}\left|m_1\left(2\cos\theta +\sin\theta\right)^2+2m_2\left(e^{i\delta}\cos\theta -\sin\theta\right)^2+3m_3\sin^2\theta\right|\,.
% \end{eqnarray}
\begin{eqnarray}
m_{ee, \alpha\beta\alpha}^{cTBM} = \frac{1}{6}\left|m_1 \left(2 \cos \theta e^{i\delta} +\sin \theta  \right)^2+ 2m_2 \left(e^{i \delta}\cos\theta -\sin\theta\right)^2+3 m_3\sin^2\theta \right|\,.
\end{eqnarray}
Imposing the current restrictions from neutrino oscillations~\cite{deSalas:2017kay} one finds nearly identical
results for the $0\nu\beta\beta$ decay amplitude parameter $m_{ee}$ also for this case. Indeed, $m_{ee}$ in the $\alpha \beta \alpha$ scenario only differs from the $\alpha \alpha \beta$ case by a slightly different value for $\theta_{23}$ (see Fig.~\ref{fig:plot13vs23}) and a slightly different correlation for $\theta_{12}$ vs $\delta_{CP}$ (see Fig.~\ref{fig:12vsdeltaCP}). The atmospheric angle is nearly maximal in both cases, but in different octant.

\section{Summary}
\label{sec:summary}

Starting from the complex version of the Tri-Bi-Maximal lepton mixing pattern we have examined the generalized CP symmetries of the charged lepton mass matrix. These symmetries are employed in order to 'revamp' the simplest TBM \textit{Ansatz} for the lepton mixing matrix in a systematic manner. The resulting generalized patterns share some of the attractive features of the original TBM matrix, while being consistent with current oscillation experiments indicating non-vanishing $\theta_{13}$. We have explicitly examined the case where two CP symmetries are preserved in the charged lepton sector, the resulting predictions given in Eqs.~\ref{eq:eq:sumR_clep_1} and \ref{eq:eq:sumR_clep_2}
and Figs.~\ref{fig:plot13vs23}-\ref{plots132}. We have also briefly discussed some of the phenomenological implications of our new mixing patterns, both for neutrino oscillation as well as neutrinoless double beta decay search experiments, illustrated in Figs.\ref{fig:oscillationT2k}, \ref{fig:oscillationNova}, \ref{fig:oscillationDune} and \ref{fig:nu0eeaab}. % For completeness we have also briefly discussed the case where only one generalized CP symmetry is preserved by the charged lepton mass matrix.
Finally, we mention that the same systematic procedure may be employed in order to 'revamp' other \textit{a priori} unrealistic patterns of neutrino mixing.
\vspace{-1cm}
\section{Acknowledgments}

This work is supported by National Natural Science Foundation of China under Grant Nos 11835013, 11522546 and 11847240 and China Postdoctoral Science Foundation under Grant Nos 2018M642700 and the Spanish grants FPA2017-85216-P (AEI/FEDER, UE), SEV-2014-0398 and PROMETEO/2018/165 (Generalitat  Valenciana). We thank the support of the Spanish Red Consolider MultiDark FPA2017-90566-REDC. S.C.C is also supported by the FPI grant BES-2016-076643.

\bibliographystyle{utphys}
\bibliography{bibliography}
\end{document}